\documentclass[twoside,letterpaper]{IEEEtran}

\makeatletter
\newif\if@restonecol
\makeatother

\usepackage{times}
\usepackage[vlined,algoruled]{algorithm2e}
\usepackage{balance}
\usepackage{multirow}
\usepackage{latexsym}
\usepackage{graphicx}
\SetKwInOut{Input}{Input}
\SetKwInOut{Output}{Output}

\SetKw{Break}{break}

\SetKw{Continue}{continue}

\SetKwFunction{Pairwise}{PairwiseDetection}
\SetKwFunction{Originals}{FindOriginals}
\SetKwFunction{Global}{GlobalPr}
\SetKwFunction{Max}{max}

\SetCommentSty{textsl}
\SetAlgoSkip{1pt}

\newcommand{\eat}[1]{}
\newcommand{\ie}{{\em i.e.}}
\newcommand{\eg}{{\em e.g.}}

\newcommand{\tightlist}{\itemsep=-3pt}

\newcommand{\rbox}{\hfill $\Box$}

\newtheorem{definition}{Definition}[section]
\newtheorem{proposition}[definition]{Proposition}

\newtheorem{example}[definition]{Example}

\begin{document}

\title{Scaling up Copy Detection}


\author{%
{Xian Li{\small $~^{1}$}, Xin Luna Dong{\small $~^{2}$}, Kenneth B. Lyons{\small $~^{3}$},
 Weiyi Meng{\small $~^{1}$}, Divesh Srivastava{\small $~^{3}$}
 }%
\vspace{1.6mm}\\
\fontsize{10}{10}\selectfont\itshape
$^{1}$\, Computer Science Department of Binghamton University, \{xianli, meng\}@cs.binghamton.edu\\
%
\fontsize{10}{10}\selectfont\itshape
$^{2}$\,Google Inc, lunadong@google.com
 
\fontsize{10}{10}\selectfont\itshape
$^{3}$\, AT\&T Labs-Research, \{kbl, divesh\}@research.att.com
}


\maketitle
\thispagestyle{empty}
\pagestyle{empty} 
\begin{abstract}
Recent research shows that copying is prevalent for Deep-Web data and 
considering copying can significantly improve truth finding from
conflicting values. However, existing copy detection techniques
do not scale for large sizes and numbers of data sources,
so truth finding can be slowed down by one to two orders of magnitude
compared with the corresponding techniques that do not consider copying.
In this paper, we study {\em how to improve scalability of copy detection 
on structured data}.

Our algorithm builds an inverted index for each \emph{shared} value
and processes the index entries in decreasing order of how much the shared value 
can contribute to the conclusion of copying. We show how we use the index
to prune the data items we consider for each pair
of sources, and to incrementally refine our results in iterative 
copy detection. We also apply a sampling strategy
with which we are able to further reduce copy-detection time while still
obtaining very similar results as on the whole data set.
Experiments on various real data sets show that 
our algorithm can reduce the time for copy detection
by two to three orders of magnitude; in other words, truth finding 
can benefit from copy detection with very little overhead.
\end{abstract}


\section{Introduction}
\label{sec:intro}
As we enjoy the abundance of information on the Web, 
we are often confused and misguided by low-quality data, which can be 
out-of-date, incomplete, or erroneous.
Recently, Li et al.~\cite{LDL+12} showed that even for domains such as 
stock and flight, conflicting values are provided by different Deep Web
sources on 70\% of the {\em data items} (\eg, closing price of a stock).
In addition, although well-known authoritative sources, such as {\em NASDAQ}
for stock and {\em Orbitz} for flight, often have fairly high accuracy, 
they may not have the desired coverage. 
Many applications, such as integrating 
Web-scale data and building knowledge bases from the Web,
call for advanced {\em data fusion} techniques to resolve conflicts 
from different sources and identify values that reflect the real world.

Typically, we expect that the values provided by many sources are likely
to be true. Unfortunately, data copying is common on the Web: 
a false value can spread through copying and become quite popular.
Thus, we need to detect copying and discount data from copiers for truth finding. 
Most of current copy-detection techniques on structured data~\cite{BCM+10, DBS09a, DBS09b, QAH+13}
share two features. First, they exploit the intuition that 
copying is likely if {\em many false values} are shared, since this is 
unlikely to happen between independent sources. Second, since 
the truthfulness of data is often unknown {\em a priori}, they iteratively conduct 
copy detection, truth finding, and source-accuracy computation
until convergence. In~\cite{LDL+12} the authors show that 
the copy-detection techniques of ~\cite{DBS09a} 
can significantly improve truth-finding results in the presence of 
copied values and fix half of the errors made by naive voting or
considering only source accuracy. 
Detecting copying between Web sources is also important to finding the truths in building knowledge bases~\cite{DGK+14}, which are widely used in Web search.
Copy detection is also valuable
in studying dissemination of
information, protecting the rights of data providers, and so on.

Despite its importance, research on copy detection 
for structured data is still in its infancy, focusing mainly on effectiveness 
of the techniques. Current techniques examine every shared data item 
between every pair of sources in each iteration to detect copying, 
so do not scale when the sizes of the sources,
the number of the sources, or the number of iterations is large.  
As pointed out in~\cite{LDL+12},
even on a medium-sized data set (55 sources and 16,000 data items),
conducting copy detection would slow down data fusion by
one to two orders of magnitude.
In the Big Data environment where the data sources are growing rapidly
in size and millions of sources in the same domain are emerging~\cite{DMP12},
scalability is important for successful copy detection.
Consequently, we propose and study the problem of {\em how to improve
scalability of copy detection for structured data.}

Before we describe our techniques, it is instructive to consider scalable techniques that have been proposed for discovering copying on other types of data, such as text documents and software programs
(surveyed in~\cite{DS11}). Each document or program can be considered as 
a text sequence. Reuse of sufficiently large text fragments
is taken as evidence for copying and copy detection essentially
looks for such fragments. To improve scalability, proposed 
techniques create {\em signatures} (or {\em fingerprints}) of each
text fragment and build an {\em inverted index} for all or selected
signatures. A pair of documents or programs are compared only if
a sufficiently large number of signatures are shared. Directly applying such 
techniques on structured data would be inadequate for two reasons. 
First, different values need to be treated
differently in copy detection: sharing false values is treated as strong
evidence for copying whereas sharing true ones is considered
as a possible coincidence and treated only as weak evidence;
thus, whether copying is likely depends not only on the number of 
shared values but also on the truthfulness of the shared values.
Second, there is no natural way to order structured data
(records and attributes); thus, large text fragments may not be shared 
by sources even if there is copying. 

We propose a comprehensive set of index structures and algorithms for this problem 
and make the following contributions. First, we design an inverted index, 
where each entry corresponds
to a shared value for a particular data item and lists the data sources
that provide this value. The presence of a source in an index entry guarantees
its absence in all entries that correspond to other values for the same data item. 
We associate each entry with a score, derived
from the probability of the value being false;
the higher the score, the stronger the evidence that sharing the entry 
can serve for detecting copying.
We process the entries in decreasing order of the scores and consider a pair
of sources only if they share at least some value with a high score
(Section~\ref{sec:index}). 

Second, we propose pruning algorithms to further improve scalability for 
copy detection. Since we process index entries in decreasing order of their scores, we consider
strong evidence early as we scan the index. Once we have accumulated 
enough evidence to decide copying or no-copying, 
we can stop without considering every shared value (Section~\ref{sec:singleround}).

Third, we develop incremental algorithms for iterative copy detection.
We observe that between consecutive iterations,
our truth-finding decisions typically change very slightly,
and so do our decisions on copying.
Instead of detecting copying from scratch in each iteration, 
we refine our decisions incrementally from previous iterations
(Section~\ref{sec:incremental}).

\eat{Fourth, we design sampling strategies for copy detection that takes into account
the skew of the number of data items in data sources to 
reduce the number of items we consider while still obtaining
very similar results as on the whole data set
(Section~\ref{sec:sample}). }

Finally, we experimented on a variety of real data sets, 
showing high scalability of the proposed techniques and big
performance gains over simple sampling strategies 
(Section~\ref{sec:experiment}). We show that
our algorithms together can speed up copy detection by
two to three orders of magnitude, and can detect copying for thousands of sources in seconds even on a single server.  With our algorithms, copy detection can significantly improve truth finding with very little overhead.

Our index and pruning techniques also shed light on 
other applications that require computing similarity by accumulating 
weighted evidence; 
for example, in record linkage different attributes may have different weights in the computation of record similarity.

\eat{
(1) using the inverted index itself can tremendously reduce
the number of source pairs we consider and reduce
copy-detection time by one to two orders of magnitude;
(2) pruning and incremental detection can significantly 
reduce the number of data items we consider for each pair of sources
and further reduce copy-detection time by nearly one order
of magnitude; and
(3) a careful sampling can additionally reduce execution time by 
orders of magnitude without sacrificing the quality of copy detection much.
}

\section{Preliminaries}
\label{sec:overview}
We review copy-detection techniques for structured data
and describe the opportunities to improve scalability.

\subsection{Copy detection}
Copy detection for structured sources was recently studied
in~\cite{BCM+10, DBH+10a, DBS09a, DBS09b, QAH+13}. They consider
a domain $\cal D$ of {\em data items}, each describing a
particular aspect of a real world object, such as the capital
of a state. They consider a set $\cal S$ of {\em data sources},
each providing data for a subset of data items in $\cal D$;
we denote by $\bar D(S)$ the items provided by $S \in \cal S$.
Schema mapping and entity resolution are assumed to have been performed so 
it is known which data items are shared between the sources 
(errors in these stages can be treated as wrongly provided data).
A source $S_1$ is considered as a {\em copier} of $S_2$ if
$S_1$ copies a subset of data values from $S_2$. {\em Copy detection}
aims at {\em finding copying between sources in $\cal S$}.

\smallskip
\noindent
{\bf Bayesian analysis:}
The key idea in copy detection is to examine the values shared between
a pair of sources.\footnote{\small Advanced techniques also consider
coverage and formatting of data items~\cite{DBH+10a}, 
for which we can extend
our techniques.} It is assumed that each data
item is associated with a single true value that reflects the
real world, but there are in addition $n>1$ false values in the domain for each data item, 
and they are uniformly distributed.\footnote{\small This assumption can
be relaxed to take value distributions into account~\cite{DBS09a},
but is used here for simplicity.} Thus, the likelihood that two
independent sources share the same false value is typically low. 
As a result, sharing false values on a large number of data items 
serves as strong evidence for copying. 

Based on this intuition, Bayesian analysis is conducted for copy 
detection~\cite{BCM+10, DBH+10a, DBS09a, DBS09b}. 
Consider two sources $S_1$ and $S_2$ and let $\Phi$ be the 
observation on their data. Denote by $S_1 \to S_2$ 
(or $S_2 \leftarrow S_1$) copying by
$S_1$ from $S_2$, and by $S_1 \bot S_2$ no-copying between 
them.\footnote{\small We can also extend our techniques to
distinguish direct copying from co-copying and transitive 
copying~\cite{DBH+10a} and we skip the details.}
It is assumed that there is no mutual copying
($S_1$ copies from $S_2$ {\em and} $S_2$ copies from $S_1$), so 

\vspace{-.15in}
{\small
\begin{eqnarray}
\nonumber
&& Pr(S_1 \bot S_2|\Phi) \\
&=& {\beta Pr(\Phi|S_1 \bot S_2)
\over \beta Pr(\Phi|S_1 \bot S_2)+\alpha Pr(\Phi|S_1 \to S_2)
      +\alpha Pr(\Phi|S_1 \leftarrow S_2)}.
\label{eqn:Bayes}
\end{eqnarray}
}
\vspace{-.1in}

\noindent
Here, $0<\alpha<.5$ is the a-priori probability of a source
copying from another one and $\beta=1-2\alpha$. 
Assuming independence between data items, and denoting
by $\Phi_D$ the observation on data item $D$, we have
$Pr(\Phi|S_1 \bot S_2)=\Pi_{D \in {\cal D}}Pr(\Phi_D|S_1 \bot S_2)$
(similar for other cases). Thus, we can rewrite
Eq.(\ref{eqn:Bayes}) as follows.

\vspace{-.15in}
{\small
\begin{eqnarray}
\label{eqn:ratio}
\nonumber
&& Pr(S_1 \bot S_2|\Phi)\\
&=& {1 \over 1+{\alpha \over \beta}(
\Pi_{D \in {\cal D}}{Pr(\Phi_D|S_1 \to S_2) \over Pr(\Phi_D|S_1 \bot S_2)}+
\Pi_{D \in {\cal D}}{Pr(\Phi_D|S_1 \leftarrow S_2) \over Pr(\Phi_D|S_1 \bot S_2)})}.
\end{eqnarray}
}
\vspace{-.1in}

Computation of Eq.(\ref{eqn:ratio}) would require computing 
$\Pi_{D \in {\cal D}}{Pr(\Phi_D|S_1 \to S_2) \over Pr(\Phi_D|S_1 \bot S_2)}$; 
we denote its logarithms by $C_\to= \sum_{D \in {\cal D}}\ln{Pr(\Phi_D|S_1 \to S_2) \over 
Pr(\Phi_D|S_1 \bot S_2)}$ (similarly, 
$C_\leftarrow=\sum_{D \in {\cal D}}\ln{Pr(\Phi_D|S_1 \leftarrow S_2) \over 
Pr(\Phi_D|S_1 \bot S_2)}$). Essentially, $C_\to$ and $C_\leftarrow$ 
accumulate the contribution from each data item $D \in \cal D$. 
We denote the {\em contribution score} from $D$ to $C_\to$ by
$C_\to(D)=\ln{Pr(\Phi_D|S_1 \to S_2) \over Pr(\Phi_D|S_1 \bot S_2)}$
and compute it as follows (similar for $C_\leftarrow$).

{\em 1. Providing the same value $v$:} 
    For each source $S \in \cal S$, its {\em accuracy},
    denoted by $A(S)$, is measured as the fraction of its true values over 
    all provided values. This can be considered
    as the probability of $S$ providing a true value
    for a data item. Then, the probability of
    $S_1$ and $S_2$ independently providing the (same) true value is 
    $A(S_1)A(S_2)$, and that for the same false value 
    is ${1-A(S_1) \over n}\cdot {1-A(S_2) \over n}
    \cdot n = {(1-A(S_1))(1-A(S_2)) \over n}$ (recall it is
    assumed that there are $n$ uniformly distributed false values). 
    In practice, we are often not sure which value is true.
    Let $P(D.v)$ be the probability of value $v$ being true for $D$, 
    then 

\vspace{-.1in}
{\small
\begin{eqnarray}
\nonumber
Pr(\Phi_D|S_1 \bot S_2) &=& P(D.v)A(S_1)A(S_2)\\
&+& (1-P(D.v)){(1-A(S_1))(1-A(S_2)) \over n}.
\label{eqn:indep}
\end{eqnarray}
}
\vspace{-.15in}

    Now consider copying.
    Let $s$ be the {\em selectivity} of copying\footnote{\small
    $\alpha , n, s$ are inputs and can be set/refined
    according to ~\cite{DBH+10a, DBS09a}.}; that is,
    the probability that the copier copies on a particular item. 
    When $S_1$ copies from $S_2$ on $D$ (with probability $s$), 
    they must provide the same value. The probability of
    our observed value then depends on the likelihood that
    $S_2$ provides the value. The probability of $S_2$
    providing the true value is $A(S_2)$ and that for
    a false value is $1-A(S_2)$. Thus, the probability for
    our observation of $S_2$'s data on $D$, denoted by $\Phi_D(S_2)$, is

\vspace{-.15in}
{\small
\begin{equation}
Pr(\Phi_D(S_2))=P(D.v)A(S_2)+(1-P(D.v))(1-A(S_2)). \label{eqn:copy}
\end{equation}
}
\vspace{-.15in}

    When $S_1$ does not copy from $S_2$ on $D$ (with probability 
    $1-s$), the probability that they both (independently) provide
    $v$ is the same as $Pr(\Phi_D|S_1 \bot S_2)$. Thus, 

\vspace{-.15in}
{\small
\begin{equation}\label{eqn:no-copy}
Pr(\Phi_D|S_1 \to S_2) = (1-s)Pr(\Phi_D|S_1 \bot S_2)+sPr(\Phi_D(S_2)).
\end{equation}
}
\vspace{-.1in}

Combining Eq.(\ref{eqn:indep}-\ref{eqn:no-copy}), we have

\vspace{-.1in}
{\small
\begin{equation}\label{eqn:samevalue}
C_\to(D) = \ln(1-s+s\cdot{Pr(\Phi_D(S_2)) \over
Pr(\Phi_D|S_1 \bot S_2)}).
\end{equation}
}
\vspace{-.1in}

{\em 2. Providing different values:} When two sources provide
    different values, the copier cannot copy (the probability is $1-s$)
and they independently provide different values. Thus, 

\vspace{-.1in}
{\small
\begin{eqnarray}
Pr(\Phi_D|S_1 \to S_2)&=&(1-s)Pr(\Phi_D|S_1 \bot S_2);\\
C_\to(D) &=& \ln(1-s). \label{eqn:diffvalue}
\end{eqnarray}
}
\vspace{-.1in}

It has been proved that $C_\to(D)$ is positive when $S_1$ and $S_2$
share the same value on $D$ and negative otherwise, and
it is larger when the shared value has a lower $P(D.v)$ 
(\ie, $v$ is more likely to be false)~\cite{DBS09a}.
In other words, sharing a value serves as evidence for copying
and vice versa, and sharing a false value serves as strong evidence
for copying. 

\begin{table}
\vspace{-.1in}
\centering
{\small
\caption{Motivating example. False values are in italic font.
An empty cell corresponds to a missing value from a source.
\label{tbl:motivate}}
\begin{tabular}{|p{0.15cm}|p{0.5cm}|p{1.05cm}|p{1.05cm}|p{1.05cm}|p{1.05cm}|p{1.05cm}|}
\hline
           & Accu &    \shortstack{NJ \\ ($D_1$)}         &  \shortstack{AZ\\ ($D_2$)}     &  \shortstack{NY \\ ($D_3$)}        &  \shortstack{FL \\($D_4$)}       &   \shortstack{ TX \\($D_5$)} \\
\hline
$S_0$   &   0.99   & Trenton &   Phoenix  &   Albany      &                      &  Austin \\
$S_1$   &   0.99   & Trenton &   Phoenix  &   Albany      &   Orlando         &  Austin \\
$S_2$   &   0.2   & {\em Atlantic} &   Phoenix  &   {\em NewYork}  &   {\em Miami}         &  {\em Houston} \\
$S_3$   &   0.2 & {\em Atlantic} &   Phoenix  &   {\em NewYork}  &   {\em Miami}         &  {\em Arlington} \\
$S_4$   &   0.4   & {\em Atlantic} &   Phoenix   &   {\em NewYork}  &  Orlando         &  {\em Houston} \\
$S_5$   &   0.6   & {\em Union}    &   {\em Tempe}  &   Albany      &   Orlando         &  Austin \\
$S_6$   &   0.01 &              &     {\em Tempe}   &   {\em Buffalo}     &   {\em PalmBay}  &  {\em Dallas} \\
$S_7$   &   0.25   & Trenton &   		 &   {\em Buffalo}     &   {\em PalmBay}         &  {\em Dallas} \\
$S_8$   &   0.2 & Trenton &   {\em Tucson} &   {\em Buffalo}     &   {\em PalmBay}         &   {\em Dallas} \\  
$S_9$ &   0.99     & Trenton &                         &                   &   Orlando         &  Austin \\
\hline
\end{tabular}}
\end{table}
\begin{example}\label{eg:motivate}
Consider 10 data sources that describe capitals for 5 states
in the US (Table~\ref{tbl:motivate}); their accuracy measures
are shown on the second column. There is copying between
$S_2-S_4$ and between $S_6-S_8$. We set $\alpha=0.1, s=0.8,$ 
and $n=50$.

Consider $S_2$ and $S_3$ as an example. Starting with $D_1$,
they provide the same value so we apply Eq.(\ref{eqn:samevalue}).
Suppose that {\sf NJ.Atlantic} has probability .01
to be true. Then, 
$C_\to(D_1)=C_\leftarrow(D_1)=\ln(.2+.8\cdot {.01*.2+.99*.8 \over .01*.2*.2+.99*{.8*.8\over 50}})= 3.89$, 
showing that sharing this false value is strong evidence for copying. 
We compute for other items similarly and eventually
$C_\to=C_\leftarrow=3.89+1.6+3.86+3.83-1.6=11.58$.
Applying Eq.(\ref{eqn:ratio}) computes $Pr(S_2 \bot S_3|\Phi)=.00004$, 
so copying is very likely.

Now consider $S_0$ and $S_1$, which also share 4 values.
However, suppose we know that these values are all true
and each of them has a contribution .01 (details skipped).
Eventually, $C_\to=C_\leftarrow=.01*4=.04$ and
$Pr(S_0 \bot S_1|\Phi)=.79$, so copying is unlikely.\rbox
\end{example}

\begin{table}
\vspace{-.1in}
\centering
{\small
\caption{\label{tbl:iteration}Iterations for the motivating example.}
\begin{tabular}{|c|c|c|c|c|c|c|c|}
\hline
	     &   Rnd 1   &   Rnd 2    &  Rnd 3   &   Rnd 4 & Rnd 5\\
\hline
$S_0$ &	0.75 &	0.94 &	0.96 &	0.98 &	0.99\\
$S_1$ &	0.98 &	0.99 &	0.99 &	0.99 &	0.99\\
$S_2$ &	0.38 &	0.23 &	0.21 &	0.2  &	0.2 \\
$S_3$ &	0.38 &	0.23 &	0.21 &	0.2  &	0.2 \\
$S_4$ &	0.58 &	0.43 &	0.41 &	0.4  &	0.4 \\
\hline
\end{tabular}

\vspace{.1in}
(a) Source accuracy.

\vspace{.1in}
\begin{tabular}{|c|c|c|c|c|c|c|c|}
\hline
	     &   Rnd 1   &   Rnd 2    &  Rnd 3   &   Rnd 4 & Rnd 5\\
\hline
NJ.Trenton &	0.9 &	0.95 &	0.96 &	0.97 &	0.97\\
NJ.Atlantic &	0.07 &	0.03 &	0.02 &	0.01 &	0.01\\
AZ.Phoenix &	0.94 &	0.95 &	0.95 &	0.95 &	0.95\\
NY.Albany &	0.07 &	0.77 &	0.88 &	0.92 &	0.94\\
NY.NewYork	&       0.84 &	0.16 &	0.08 &	0.03 &	0.02\\
FL.Orlando &	0.9 &	0.92 &	0.92 &	0.92 &	0.92\\
FL.Miami & 	0.05 &	0.03 &	0.04 &	0.03 &	0.03\\
TX.Austin &	0.9 &	0.93 &	0.95 &	0.96 &	0.96\\
TX.Houston &	0.04 &	0.03 &	0.02 &	0.02 &	0.02\\
\hline

\end{tabular}

\vspace{.1in}
(b) Probability of values in the index.}
\vspace{-.1in}
\end{table}

\smallskip
\noindent
{\bf Iterative computation:} 
We often do not know value probability $P(D.v)$ and source accuracy $A(S)$,
and computing them often requires knowledge of the copying
relationship (details in~\cite{DBS09a}). An iterative approach has
been proposed as follows~\cite{BCM+10, DBS09a}: 
starting with assuming the same accuracy
for each source, each round iteratively computes copying probability, 
value truthfulness, and source accuracy, until convergence.
For our motivating example, there are five rounds
before convergence. Table~\ref{tbl:iteration} shows 
the source accuracy and value probability computed in each round
(for simplicity, we show only for the first 5 sources and their values).

\subsection{Opportunities for scalability improvement}
\label{sec:opportunity}
Previous works~\cite{DBH+10a, DBS09a} conduct copy detection in 
an exhaustive fashion. For each pair of sources, the algorithm,
called {\sc Pairwise}, does the following: (1) compute for each shared $D \in \cal D$
the contribution scores $C_\to(D)$ and $C_\leftarrow(D)$;
(2) accumulate the scores and compute $C_\to$ and $C_\leftarrow$;
(3) apply Eq.(\ref{eqn:ratio}) for probability computation.
This process is repeated in every round.
If the results converge in $l$ rounds, the time complexity is
$O(l|{\cal D}||{\cal S}|^2)$. {\sc Pairwise} is not scalable if the number of sources 
or data items is large, or there are many iterations. 
The algorithm proposed in ~\cite{BCM+10} 
also examines every pair of sources, so has similar complexity to 
{\sc Pairwise}.

There are several opportunities for improving scalability
of copy detection. First, for some pairs of sources
that share no value at all or just a few true values, 
we can determine that they are independent without going through 
all of the shared data items; this can reduce the number
of source pairs we examine. For our motivating example, 
{\sc Pairwise} requires examining 45 pairs of sources;
however, among them 18 pairs (such as $S_0$ and $S_6$)
do not share any value, and the pair of $S_0$ and $S_5$ share only 
two true values. We can skip these pairs. 
Section~\ref{sec:index} describes how we can explore this 
opportunity by building and using a specialized inverted index. 

Second, for some pairs of sources
that share a lot of false values, we can determine 
copying after we observe only a subset of these false values;
this can reduce the number of data items we examine
for those pairs. In our motivating example, $S_2$ and $S_3$
share 4 values, including 3 false ones; actually, after observing 2 
false values, we can already determine copying
without knowing the rest of the provided values.
Section~\ref{sec:singleround} explores this opportunity
for single-round copy detection.

Third, in the iterative process,
the changes in value probability and source accuracy 
between two consecutive rounds after the second round
are typically very small. Thus, we can do
copy detection incrementally to consider fewer data items for each
pair of sources in later rounds. 
Section~\ref{sec:incremental} explores this opportunity
and describes incremental copy detection.

\eat{
Fourth, we can sample a subset
of data items from $\cal D$ for copy detection; 
this can also reduce the number of data items we consider.
Section~\ref{sec:sample} describes our sampling approach.

Finally, recent work has extended the basic copy-detection 
technique we reviewed by considering source coverage,
value format, different copying patterns, and global 
analysis~\cite{BCM+10, DBH+10a}.
Section~\ref{sec:extend} extends our techniques for advanced
copy detection. 
}

Ideally, these aforementioned optimizations should tremendously reduce
computation and thus execution time, while leading to the same
(binary) decision on copying relationships, and also on 
value truthfulness. In practice, however, early pruning 
may improve efficiency with a slight loss of accuracy.
We show in experiments (Section~\ref{sec:experiment})
the effectiveness and scalability of our techniques. 

We have also explored Fagin's NRA (No Random Access) algorithm ~\cite{FLN01} for top-$k$ search to speed up copy detection.
We maintain for each value of a data item
a list of contribution scores for the pairs of sources
that share the value, and order the pairs in decreasing score order.
We also maintain a list containing the accumulated 
contribution scores from different values 
for pairs of sources that have such differences.
Then, $C_\to$ (similar for $C_\leftarrow$) for a particular 
pair of sources is the sum of the scores from all lists. 
To find copying, we can apply NRA to find the pairs with top
values of $C_\to$ and $C_\leftarrow$ and stop when $C_\to$ and $C_\leftarrow$
lead to the conclusion of no-copying. 
However, we show in experiments (Section~\ref{sec:experiment}) that
even generating the input to NRA (\ie, the ordered lists) 
for our problem is slower than our proposed approaches.

\section{Inverted Index}
\label{sec:index}
We first describe an important building block in our solution--the
{\em inverted index}, which facilitates the exploration of many 
aforementioned opportunities for scalability improvement.
Inverted indexes were originally used in Information Retrieval~\cite{MRS08}
and we describe the adaptation for copy detection.

\smallskip
\noindent
{\bf Building the index:} 
An important component in copy detection is to find for each
pair of sources the values, not just the items, they share. We can facilitate 
this process with an inverted index, 
where each entry corresponds to a value $v$ for a data item
$D$, denoted by $D.v$, and contains the sources that provide $v$ on $D$. 
Note that the presence of source $S$ in the entry for $D.v$ guarantees 
that $S$ is not present in any of the entries for $D.v', v' \neq v$.

Intuitively, we wish to first consider sharing of values that serve as strong 
evidence for copying, as it provides the opportunity to prune weak evidence
for copying. We order the entries according to 
their contribution scores to $C_\to$ and $C_\leftarrow$.
Note, however, that according to Eq.(\ref{eqn:indep}-\ref{eqn:samevalue})
the contribution from sharing $D.v$ 
can be different for different pairs of sources with various accuracy; 
we choose the maximum one, denoted by $\hat M(D.v)$. 
The next proposition shows that we can compute $\hat M(D.v)$
only from providers (\ie, sources) with the maximum or minimum accuracy
(proofs omitted to save space).

\begin{proposition}
Let $D.v$ be a value with probability $P(D.v)$. Let
$A_{min}$ be the minimum accuracy among $D.v$'s providers.
\begin{itemize}
  \item If $A_{min}\leq{1 \over 1+{nP(D.v) \over 1-P(D.v)}}$,
    $\hat M(D.v)$ is obtained by Eq.(\ref{eqn:samevalue})
    when $S_1$ has the maximum accuracy
    and $S_2$ has the minimum accuracy;
  \item If $A_{min}>{1 \over 1+{nP(D.v) \over 1-P(D.v)}}$ and $P(D.v)<.5$,
    $\hat M(D.v)$ is obtained by Eq.(\ref{eqn:samevalue})
    when $S_2$ has the minimum accuracy
    and $S_1$ has the second minimum accuracy;
  \item Else,
    $\hat M(D.v)$ is obtained when $S_1$ has the minimum accuracy
    and $S_2$ has the second minimum accuracy.\rbox
\end{itemize}
\end{proposition}
\eat{\begin{proof}
According to Eq.(\ref{eqn:indep}-\ref{eqn:samevalue}), 
we examine $X={1+PA \over A(S_1)+PA{1-A(S_1) \over n}}$, 
where $P={1 \over P(D.v)}-1$ and $A={1 \over A(S_2)}-1$. 
We can prove that (1) when $A(S_2)$ decreases, both $A$ 
and $X$ increase; and (2) when $A(S_1)$ increases, $X$ increases
when $PA>n$ and decreases when $PA<n$. These lead to the results.
\end{proof}}

We can now formally define our specialized inverted index.  
\begin{definition}[Inverted Index]
Let $\cal D$ be a set of data items and $\cal S$ be a set of sources.
The {\em inverted index} for $\cal D$ and $\cal S$ contains
a set $\bf E$ of entries, such that for each $E \in {\bf E}$,
\begin{enumerate}
  \item $E$ corresponds to a value $D_E.v_E$, where $D_E \in \cal D$ 
and $v_E$ is a value provided by at least two sources on $D_E$;
  \item $E$ is associated with probability $P(E)$ for 
$D_E.v_E$ being true and with {\em contribution score} 
$C(E)=\hat M(D_E.v_E)$;
  \item the entry contains a set $\bar S(E)$ of sources that provide $D_E.v_E$.\rbox
\end{enumerate}
\end{definition}

\begin{table}
\vspace{-.1in}
\centering
{\small
\caption{Inverted index for the motivating example.
The two sources used to compute the contribution scores are in bold.
\label{tbl:inverted}}
\begin{tabular}{c|c|c|l}
\hline
Value & Pr & Score & Providers \\
\hline
AZ.Tempe & 0.02 & 4.59 & ${\bf S_5, S_6}$ \\
NJ.Atlantic & 0.01 & 4.12 & $S_2, {\bf S_3, S_4}$ \\
TX.Houston & 0.02 & 4.05 & ${\bf S_2, S_4}$ \\
NY.NewYork & 0.02 & 4.05 & $S_2, {\bf S_3, S_4}$ \\
TX.Dallas & 0.02 & 3.98 & ${\bf S_6, S_7,} S_8$ \\
NY.Buffalo & 0.04 & 3.97 & ${\bf S_6, S_7,} S_8$ \\
FL.PalmBay & 0.05 & 3.97 & ${\bf S_6, S_7,} S_8$ \\
FL.Miami & 0.03 & 3.83 & ${\bf S_2, S_3}$ \\
AZ.Phoenix & 0.95 & 1.62 & $S_0, S_1, {\bf S_2, S_3,} S_4$ \\
NJ.Trenton & 0.97 & 1.51 & $S_0, S_1, {\bf S_7, S_8,} S_9$ \\
FL.Orlando & 0.92 & 0.84 & $S_1, {\bf S_4, S_5,} S_9$ \\
NY.Albany & 0.94 & 0.43 & $S_0, {\bf S_1, S_5}$ \\
TX.Austin & 0.96 & 0.43 & $S_0, S_1, {\bf S_5, S_9}$ \\
\hline
\end{tabular}}
\vspace{-.1in}
\end{table}
\begin{example}\label{eg:baseline}
Continue with Ex.\ref{eg:motivate}. 
Table~\ref{tbl:inverted} shows the inverted index for the data,
assuming knowledge of value probability.
As an example, entry {\sf NJ.Atlantic} has probability 0.01
and contribution score 4.12, 
computed from pair $(S_4, S_3)$, with the highest and lowest accuracy
among providers of {\sf NJ.Atlantic}. 
Note that there is no entry for value {\sf NJ.Union},
{\sf AZ.Tucson}, or {\sf TX.Arlington}, 
as each of them is provided by a single source.
Also note that for any entries for the same data item,
such as {\sf NJ.Atlantic} and {\sf NJ.Trenton},
there is no overlap between their sources.
\end{example}

The following properties show that processing the entries 
in decreasing order of their contribution scores not only helps quickly accumulate strong evidence
for copying, but also helps compute the upper bound of
the contribution scores, making it amenable to 
additional optimizations. We also show in experiments 
(Section~\ref{sec:singleRoundExp}) that this processing order
significantly improves over random ordering. 
\begin{proposition}\label{prop:property}
For each pair of sources $S_1, S_2 \in \cal S$ and index entry
$E \in \bf E$, the following properties hold for $C_{\to/\leftarrow}(D_E)$.
\begin{itemize}
  \item If $S_1, S_2 \in \bar S(E)$, 
  $C_\to(D_E)$ is computed based on $P(D_E.v_E)$.
  \item If $S_1 \in \bar S(E), S_2 \not\in \bar S(E)$,
  but they share item $D_E$, they provide different
  values on $D_E$ and $C_\to(D_E)=\ln(1-s)$.
  \item If neither $S_1$ nor $S_2$ has appeared in any entry
  for $D_E$ before entry $E$, $C_\to(D_E)\leq C(E)$. \rbox
\end{itemize}
\end{proposition}

\smallskip
\noindent
{\bf Optimizing with the index:} 
With the inverted index, we can improve copy detection in three ways. 
First, copying is unlikely if two sources do not share any value;
thus, we can skip source pairs that do not appear in the same entry.

Second, copying is also unlikely if two sources
share only a few true values and we can skip them too.
To simplify the computation, we consider the entries with
the lowest contribution scores and denote by $\bar E \subseteq \bf E$ 
the subset of entries where
$\sum_{E \in \bar E}C(E)<\ln{\beta \over 2\alpha}$.
Then, for source pairs that do not share any value outside $\bar E$,
$C_\to < \ln{\beta \over 2\alpha}$ and 
$C_\leftarrow < \ln{\beta \over 2\alpha}$,
so $Pr(S_1 \bot S_2 | \Phi) > {1 \over 1+{\alpha \over \beta}({\beta \over 2\alpha}+{\beta \over 2\alpha})}= .5$ and copying is unlikely.
Thus, we consider a pair of sources only if they appear together
in some entry outside $\bar E$. 

Third, since each data item for which the two sources provide
different values contributes the same negative score $\ln(1-s)$
(Eq.(\ref{eqn:diffvalue})), the accumulated score from these items
depends only on the number of these items. This number can be derived
from (1) the number of shared items, denoted by $l(S_1,S_2)$, 
counted at index building time 
(we can apply techniques for set similarity joins~\cite{AGK06} 
to improve efficiency of counting), and (2) the number of shared
values, denoted by $n(S_1,S_2)$, counted at index scanning time.

We next describe an algorithm, {\sc Index},
that uses the inverted index for copy detection.
Instead of considering each pair of sources, {\sc Index} scans 
the inverted index in decreasing order of contribution scores and proceeds in three steps.
\begin{enumerate}
  \item For each entry $E \in {\bf E}\setminus\bar E$ and
each pair of sources $S_1, S_2 \in \bar S(E)$, 
(1) compute the contribution 
from $E$ and update $C_\to$ and $C_\leftarrow$ for $(S_1,S_2)$,
and (2) maintain $n(S_1,S_2)$.
  \item For each entry $E \in \bar E$, do the same as in Step 1 
but only for pairs encountered before.
  \item After scanning the whole index, for each already
considered pair $(S_1,S_2)$, (1) update scores for 
data items where different values are provided 
by adding $\ln(1-s)(l(S_1,S_2)-n(S_1,S_2))$,
and (2) compute copy probability accordingly.
\end{enumerate}

\begin{proposition}
Let $r$ be the number of source pairs for which we maintain scores.
{\sc Index} takes time $O(r\cdot |{\cal D}|)$ and space $O(r)$,
obtaining the same binary results as {\sc Pairwise}. Note that index building has a much lower complexity: $O(|{\cal S}||{\cal D}|)$. \rbox
\end{proposition}

The next example shows that the {\sc Index} algorithm can 
considerably improve the efficiency of copy detection.

\begin{example}\label{eg:baseline}
Continue with the motivating example. For {\sc Index}, 
the last two entries in the index (Table~\ref{tbl:inverted}) 
form the set $\bar E$ ($.43+.43<\ln{.8 \over .2}=1.39$). 
There are only 26 pairs of sources that occur in entries 
outside $\bar E$; for example, $S_0$ and $S_5$ share only
values in $\bar E$, so we do not need to consider this pair.
In total {\sc Index} needs to examine 
51 shared values and have $51*2+26*2=154$ computations
(2 additional computations for each pair of sources on different values)
for copy detection. Note that pairwise detection requires examining
45 pairs of sources and 183 shared data items, so in total
conducting $183*2=366$ computations. For this example, 
{\sc Index} cuts computation by more than half. \rbox
\end{example}

\section{Detection in One Round}
\label{sec:singleround}
{\sc Index} does not need to consider 
every pair of sources and thus can save computation;
however, for each pair it considers, it still examines all
shared values. The properties of the inverted index
(Proposition~\ref{prop:property}) make it possible to 
terminate after we examine 
only a subset of shared values for a pair. 
First, when we observe a lot of high-score (low-probability) entries
to which both sources belong, 
we may conclude with copying early.
Second, when we observe a lot of entries to which one of the two
sources belongs and a lot of high-score entries to which 
neither source belongs, we may conclude with no-copying early.
This section describes how we can speed up copy detection by 
making early decisions.

\subsection{Reducing examined shared values}
Given a pair of sources $S_1$ and $S_2$, as we scan the index,
we can maintain for $C_\to$ a maximum and a minimum score,
denoted by $C_\to^{max}$ and $C_\to^{min}$ respectively; similarly 
we maintain $C_\leftarrow^{max}$ and $C_\leftarrow^{min}$. 
If the minimum scores are large enough to conclude copying,
or the maximum scores are small enough to conclude no-copying,
we can terminate early. For such pruning, we need to (1) decide
the termination conditions and (2) compute maximum and minimum scores.

\smallskip
\noindent
{\bf Termination conditions:} We first consider binary decisions 
for copying. According to Eq.(\ref{eqn:ratio}), 
to guarantee $Pr(S_1 \bot S_2|\Phi)>.5$ (no-copying),
we should have $1 > {\alpha \over \beta}(e^{C_\to}+e^{C_\leftarrow})$;
this must be true if $C_\to^{max} < \ln{\beta \over 2\alpha}$ and
$C_\leftarrow^{max} < \ln{\beta \over 2\alpha}$. Thus, we define 
threshold $\theta_{ind}=\ln{\beta \over 2\alpha}$ for no-copying.
On the other hand, to guarantee $Pr(S_1 \bot S_2|\Phi)\leq .5$ (copying),
we should have $1 \leq {\alpha \over \beta}(e^{C_\to}+e^{C_\leftarrow})$;
this must be true if $C_\to^{min} \geq \ln{\beta \over \alpha}$
or $C_\leftarrow^{min} \geq \ln{\beta \over \alpha}$. Thus, we define
threshold $\theta_{cp}=\ln{\beta \over \alpha}$ for copying.
If none of the conditions is satisfied after we scan the whole
index, we apply Eq.(\ref{eqn:ratio}) to compute the 
probability of copying. 

If we instead wish to compute real copying probabilities
when it is between $[.1,.9]$ (or some other values close to 0 or 1),
we can consider three different cases: $Pr(S_1 \bot S_2|\Phi)>.9$,
$Pr(S_1 \bot S_2|\Phi)<.1$ 
and otherwise; we can compute the thresholds accordingly.

\smallskip
\noindent
{\bf Maximum/Minimum score computation:} 
When we scan each entry $E$, we update $C_\to^{min}$ and $C_\to^{max}$ 
for every pair $S_1, S_2 \in \bar S(E)$ as follows (similar for $C_\leftarrow$).
First, $C_\to^{min}$ is obtained when the two sources share only the
observed common values and no other value. Let
$C^0_\to(S_1, S_2)$ be the sum of scores from observed common values. 
The score for each of the remaining items is negative, $\ln(1-s)$. 
Let $n_0(S_1, S_2)$ be the number of observed shared values and
recall that $l(S_1,S_2)$ denote the number of shared items. Then,

\vspace{-.1in}
{\small
\begin{equation}
C_\to^{min}(S_1,S_2)=C^0_\to(S_1,S_2)+(l(S_1,S_2)-n_0(S_1,S_2))\ln(1-s).
\label{eqn:minscore}
\end{equation}
}
\vspace{-.1in}

For $C_\to^{max}$, we need to consider 
the already scanned entries containing only $S_1$ or $S_2$, or 
neither of them. We thus compute scores for three subsets of
data items and $C_\to^{max}$ is the sum of their scores.
\begin{itemize}\tightlist
  \item {\em Data items with observed shared values:}
    The accumulated score from such items is $C^0_\to(S_1,S_2)$.
  \item {\em Data items with observed non-shared values:}
    According to Proposition~\ref{prop:property},
    for each data item $D$ shared between $S_1$ and $S_2$,
    if we have seen only $S_1$ or $S_2$ appearing in one of
    its entries, the contribution score for $D$ is negative,
    $\ln(1-s)$. However, 
    finding the precise number of such items requires 
    recording the set of observed entries for each source
    and can cost a lot of space; thus, we estimate the minimum
    number from the numbers of observed values for $S_1$ and
    for $S_2$, denoted by $n(S_1)$ and $n(S_2)$ respectively. 
    Let $\bar N_1$ be the overlapping items among the $n(S_1)$ items 
    scanned for $S_1$--the size of $\bar N_1$ is
    roughly $n(S_1)\cdot {l(S_1, S_2) \over |\bar D(S_1)|}$;
    similar for $S_2$, denoted by $\bar N_2$. 
    The set of overlapping items among all scanned items is $\bar N_1 \cup \bar N_2$,
    their size satisfies $|\bar N_1 \cup \bar N_2| \geq \max\{|\bar N_1|, |\bar N_2|\}
    =\max\{n(S_1)\cdot {l(S_1, S_2) \over |\bar D(S_1)|},
             n(S_2)\cdot {l(S_1, S_2) \over |\bar D(S_2)|}\}$,
             denoted by $h$. Thus, the two sources provide
    different values for at least $h-n_0(S_1,S_2)$ data items,
    and the maximum score is $(h-n_0(S_1,S_2))\ln(1-s)$.
  \item {\em Data items we have not seen for $S_1$ or $S_2$:} 
    There are at most $l(S_1,S_2)-h$ such items,
    and according to Proposition~\ref{prop:property},  
    the maximum score for each of them is the score of the next unscanned
    entry, denoted by $M$. The maximum score for this 
    subset is thus $(l(S_1,S_2)-h)\cdot M$.
\end{itemize}

In summary, we have

\vspace{-.15in}
{\small
\begin{eqnarray}
\nonumber
&& C_\to^{max}(S_1,S_2) = C^0_\to(S_1,S_2)\\
&+& (h-n_0(S_1,S_2))\ln(1-s)+(l(S_1,S_2)-h)\cdot M.
\label{eqn:maxscore}
\end{eqnarray}
}
\vspace{-.15in}

\noindent
{\bf Algorithm and analysis:} From the previous analysis,
we design algorithm {\sc Bound}, which proceeds in four steps.

{\em Step I.} Build the inverted index and initialize $l(S_1,S_2)$ for each 
pair of sources that occur together in at least one entry of the index.
Initialize the active set of source pairs as ${\bf Q} = \emptyset$.

{\em Step II.} As we scan each entry $E \in {\bf E}\setminus\bar E$, 
do the following.
\begin{enumerate}\tightlist
  \item For each $S \in \bar S(E)$, if it is observed for the first time,
    set $n(S)=1$; otherwise, increase $n(S)$ by 1.
  \item For each pair $S_1, S_2 \in \bar S(E)$ that we observe for 
    the first time, set $n_0(S_1, S_2)= C^0_{\to/\leftarrow}(S_1,S_2)=0$
    and add it to $\bf Q$. 
  \item For each pair $S_1, S_2 \in \bar S(E) \cap \bf Q$, do the following.

    (1) Increase $n_0(S_1, S_2)$ by 1
    and update $C^0_{\to/\leftarrow}(S_1,S_2)$.

    (2) Compute $C_{\to/\leftarrow}^{min}(S_1,S_2)$.
    If either is above $\theta_{cp}$, 
    conclude copying and remove the pair from $\bf Q$.

    (3) Compute $C_{\to/\leftarrow}^{max}(S_1,S_2)$.
    If both are below $\theta_{ind}$, conclude no-copying and 
    remove the pair from $\bf Q$.
\end{enumerate}

{\em Step III.} As we scan each entry $E \in \bar E$, do 1 and 3 
in Step II (but only for pairs encountered before).

{\em Step IV.} After we scan the index, for each pair $(S_1,S_2) \in \bf Q$,
we have $n_0(S_1,S_2)=n(S_1,S_2)$, so $C_\to=C_\to^{min}$ (similar for $C_\leftarrow$).
If both $C_\to$ and $C_\leftarrow$ are below $\theta_{ind}$, 
conclude no-copying; otherwise, apply Eq.(\ref{eqn:ratio}).

\begin{proposition}
Let $r$ be the number of source pairs that share values,
and $e$ be the maximum number of shared entries we process 
for each pair before concluding. {\sc Bound} takes time $O(r\cdot e)$ 
and space $O(r+|{\cal S}|)$. \rbox
\end{proposition}

As {\sc Bound} estimates the number of observed overlapping data items ($h$) 
in computing $C_\to^{max}$, the result may be different from pairwise detection.
However, the computation of $h$ and the use of $M$
in Eq.(\ref{eqn:maxscore}) make the upper bounds already loose,
so the decisions are rarely different, as we observed in our experiments
(Section~\ref{sec:experiment}). 
Note that $e$ can be much smaller than $|{\cal D}|$,
so {\sc Bound} often significantly reduces
the total number of data items we consider in copy detection.
However, computing upper and lower bounds of contribution scores
introduces an overhead, so {\sc Bound} may not always 
save computation for each pair of sources, as illustrated next.

\begin{example}\label{eg:bound}
Continue with Ex.\ref{eg:baseline}. We have
$\theta_{cp}=\ln{.8 \over .1}=2.08$ and $\theta_{ind}=\ln{.8 \over .2}=1.39$.

First consider pair $(S_2, S_3)$; recall that they share
4 values (including 3 false ones) and copying is likely.
We see them first at entry {\sf NJ.Atlantic} where 
$C^0_{\to/\leftarrow}(S_2,S_3)=3.89$. 
By Eq.(\ref{eqn:minscore}) we compute 
$C_{\to/\leftarrow}^{min}=3.89-1.6*(5-1)=-2.51$.
For maximum scores, $h=1$ so by Eq.(\ref{eqn:maxscore})
$C_{\to/\leftarrow}^{max}=3.89+0+4.05*(5-1)=20.09$.
We see this pair again at entry {\sf NY.NewYork}.
We update $C^0_\to(S_2,S_3)=3.89+3.86=7.75$, so
$C_\to^{min}=7.75-1.6*(5-2)=2.95 > 2.08= \theta_{cp}$
and we can conclude copying for the pair. While {\sc Index} considers 
4 shared values for them and conducts $4*2+2=10$ computations, 
{\sc Bound} considers only 2 shared values and conducts $4+1=5$ computations.

Now consider pair $(S_0, S_1)$; recall that they share 4 true values
and no-copying is likely. When we see them at the third
shared entry {\sf NY.Albany}, we have
$C^0_{\to/\leftarrow}(S_2,S_3)=.01*3=.03$, so
$C_{\to/\leftarrow}^{max}=.03+0+0.43*(4-3)=.46<1.39=\theta_{ind}$;
we can then conclude no-copying.
Thus, {\sc Bound} considers 3 shared values and 
conducts $4*3=12$ computations.
However, {\sc Index} considers 4 shared values for them 
but conducts only $4*2+2=10$ computations, fewer than {\sc Bound}.

In total {\sc Bound} considers 26 pairs, 33 shared values,
and requires 116 computations.
It considers 18 fewer shared values and conducts 38 fewer 
computations than {\sc Index}. \rbox
\end{example}

\subsection{Reducing computation}
Although {\sc Bound} reduces the number of shared values we
consider, it introduces the overhead of computing
$C^{min}_{\to/\leftarrow}$ and $C^{max}_{\to/\leftarrow}$.
Actually, we do not need to maintain $C^{min}_{\to/\leftarrow}$
and $C^{max}_{\to/\leftarrow}$ each time we scan a shared entry;
we only need to do so when termination is likely.
This can further reduce computation for {\sc Bound}.

First, suppose after scanning entry $E$, we compute 
$C_\to^{min} < \theta_{cp}$ and $C_\leftarrow^{min} < \theta_{cp}$
for source pair $(S_1,S_2)$. The next shared value can increase
$C_\to^{min}$ and $C_\leftarrow^{min}$ by at most $M-\ln(1-s)$
(recall that $M$ denotes the contribution score of the next entry).
Thus, we do not need to re-compute $C_{\to/\leftarrow}^{min}$ until we have observed
at least $T^{min}=\lceil {\theta_{cp}-\max\{C_\to^{min},C_\leftarrow^{min}\} 
\over M-\ln(1-s)} \rceil$ shared values. 

Similarly, suppose after we scan $E$, we compute 
$C_\to^{max} \geq \theta_{ind}$ or $C_\leftarrow^{max} \geq \theta_{ind}$
for sources $(S_1,S_2)$. A new data item on which
$S_1$ and $S_2$ provide different values would reduce 
$C_\to^{max}$ and $C_\leftarrow^{max}$ by $M-\ln(1-s)$, 
so we would resume computing maximum scores when we see
$T^{max}_0=\lceil {\max\{C_\to^{max},C_\leftarrow^{max}\}-\theta_{ind} 
\over M-\ln(1-s)} \rceil$ more different values.
At entry $E$ we have already seen $h-n_0(S_1,S_2)$ different
values, so we need to see in total $T^{max}_0+h-n_0(S_1,S_2)$ 
different values. We do not re-compute $C_{\to/\leftarrow}^{max}$ 
until $n(S_1)\geq T_1^{max}=\lceil (T^{max}_0+h-n_0(S_1,S_2))\cdot{|\bar D(S_1)| \over l(S_1,S_2)} \rceil$
or $n(S_2) \geq T_2^{max}=\lceil (T^{max}_0+h-n_0(S_1,S_2))\cdot{|\bar D(S_2)| \over l(S_1,S_2)} \rceil$.

\eat{
entries for $S_2$. Thus, when we examine a shared value later, 
we do not re-compute $C_{\to/\leftarrow}^{max}$ until 
$n(S_1)\geq T_1^{max}$ or $n(S_2)\geq T_2^{max}$. 
Note that $C_\to^{max}$ and $C_\leftarrow^{max}$ are also reducing
as we scan the index since contributions from the data items
that we have not seen for $S_1$ or $S_2$ are also diminishing;
thus, unlike for the lower bound,
$C_\to^{max}$ and $C_\leftarrow^{max}$ may hit $\theta_{ind}$ before
we check them next time and so we may check more shared values
than necessary. However, this would
not affect the conclusion, as we observe the same termination
condition.
}

\begin{example}
Consider a pair of sources $S_1$ and $S_2$ that share 101 data items. 
Suppose again that $\ln(1-s)=-1.6, \theta_{cp}=2.08, \theta_{ind}=1.39$. 
Suppose the first shared item we have observed
between $S_1$ and $S_2$ has contribution $C^0_{\to/\leftarrow}=5$. 
Then $C_{\to/\leftarrow}^{min}=5-(101-1)*1.6=-155<2.08$. 
Suppose $M=4$, then 
$T^{min}=\lceil {2.08-(-155) \over 4-(-1.6)} \rceil=29$,
so we do not compute $C_{\to/\leftarrow}^{min}$ until we have observed
29 other shared values. 

For maximum scores, suppose we have not seen other entries containing
$S_1$ or $S_2$ yet, so $h=1$ and $C_{\to/\leftarrow}^{max}=5+0+(101-1)*4=405$. 
Then, $T^{max}_0=\lceil {405-1.39 \over 4-(-1.6)} \rceil =72$. Suppose 
${l(S_1,S_2) \over |\bar D(S_1)|}={l(S_1,S_2) \over |\bar D(S_2)|}=.8$,
then $T_1^{max}=T_2^{max}=\lceil {(72+1-1)\over .8} \rceil=90$. So we do not
compute $C_{\to/\leftarrow}^{max}$ until $n(S_1)\geq 90$ or
$n(S_2) \geq 90$. \rbox
\end{example}

\eat{
Second, for sources that share only true values, we believe
copying is less likely to happen. Thus, while we typically compute
$C_{\to/\leftarrow}^{min}$ first, starting from the first entry whose probability is
high (above a threshold, which we set to .9 as such values are 
typically true), we first compute $C_{\to/\leftarrow}^{max}$ for
pairs of sources we have not encountered before, so can save
computation for $C_{\to/\leftarrow}^{min}$ at termination.
}

We can improve {\sc Bound} accordingly
and the result is called {\sc Bound+}. Note that {\sc Bound+}
has the same asymptotic complexity as {\sc Bound} but in practice
can save a lot of computation. 

\eat{The next example illustrates
the advantage of {\sc Bound+} over {\sc Bound}.

\begin{example}
Continue with Example~\ref{eg:bound} and consider $(S_0, S_1)$. 
The first time we observe it is at entry {\sf AZ.Phoenix}, whose
probability is $.95>.9$. So we compute $C_{\to,\leftarrow}^{max}$ before
$C_{\to,\leftarrow}^{min}$. We conclude with no-copying for this pair after we observe
the next entry, so save computations for $C_{\to/\leftarrow}^{min}$.
The computation is thus $4+2=6<8$. For this example,
while {\sc Bound+} examines the same number of source pairs and 
shared values, it conducts only 76 computations. In addition, 
it conducts 24 computations for computing $T^{min}$ and $T_{1/2}^{max}$, 
so in total conducts 110 computations, saving 16 computations
from {\sc Bound}. \rbox
\end{example}
}

Finally, intuitively only when two sources share a lot of data
items, we are likely to significantly reduce the number of considered
shared values and compensate for the extra cost for bound computation.
We can thus apply {\sc Index} for pairs of sources that share only 
a few data items and apply {\sc Bound+} for the rest of the pairs. 
We call the resulting algorithm {\sc Hybrid}. Our experiments 
(Section~\ref{sec:experiment}) show that {\sc Hybrid} can further 
reduce computation and copy-detection time.

\eat{
\begin{example}
For the motivating example, we apply {\sc Bound+} for pairs of sources
that share at least 4 data items and apply {\sc Index} otherwise.
There are ??? pairs that share no more than 4 data items and {\sc Index}
conducts ??? less computations than {\sc Bound+} on these data items.
So in total {\sc Hybrid} conducts 69?? computations, saving 55\%
computations compared with {\sc Index} and 81\% computations compared 
with {\sc Pairwise}. \rbox
\end{example}
}

\section{Incremental Detection}
\label{sec:incremental}
Section~\ref{sec:singleround} considered copy detection 
in one single round; this section considers the iterative process.
Our observation is that although there can be changes 
on value probability and source accuracy from 
round to round, after the second round the changes are typically
small and seldom change our copy-detection decisions.
A natural thought for improving scalability is to detect 
copying incrementally after the second round.
We base our discussions on the {\sc Hybrid} algorithm.

\subsection{Overview}
Both changes in value probability $P(D.v)$ and changes in source accuracy $A(S)$
can affect copy detection. We distinguish big changes and small changes.
If a pair of sources contains a source with big accuracy change,
we need to recompute the probability of copying. For the rest of 
the pairs, we can incrementally update the contribution scores.
We {\em update scores on big-change entries first;
only for pairs whose score changes can lead to an opposite decision on copying,
we would further consider small-change entries.} 
The challenge is to reduce the number of entries we consider whenever possible
but still reach the same copying decision. 

We denote by $P_{old}(D.v)$ the probability used previously
in score contribution. Note that the recorded probability may not be 
the one from the previous round, but can be from some earlier round 
when we do the last re-computation.
For an entry $D.v$, we consider the change on $\hat M(D.v)$
rather than on $P(D.v)$, since a small change of the latter
may cause a big change of the former, which eventually
matters in score computation. To separate the change of value 
probability from that of source accuracy, we compute $\hat M(D.v)$ 
on the same two accuracies used in the round with $P_{old}(D.v)$.
Finally, to classify big and small changes, we have a threshold $\rho$.
We can either set a default one, or order the changes in decreasing
order, choose the maximum gap between two consecutive changes, and
set $\rho$ to the change above the gap. 

\begin{figure}[t]
\includegraphics[width=3.6in]{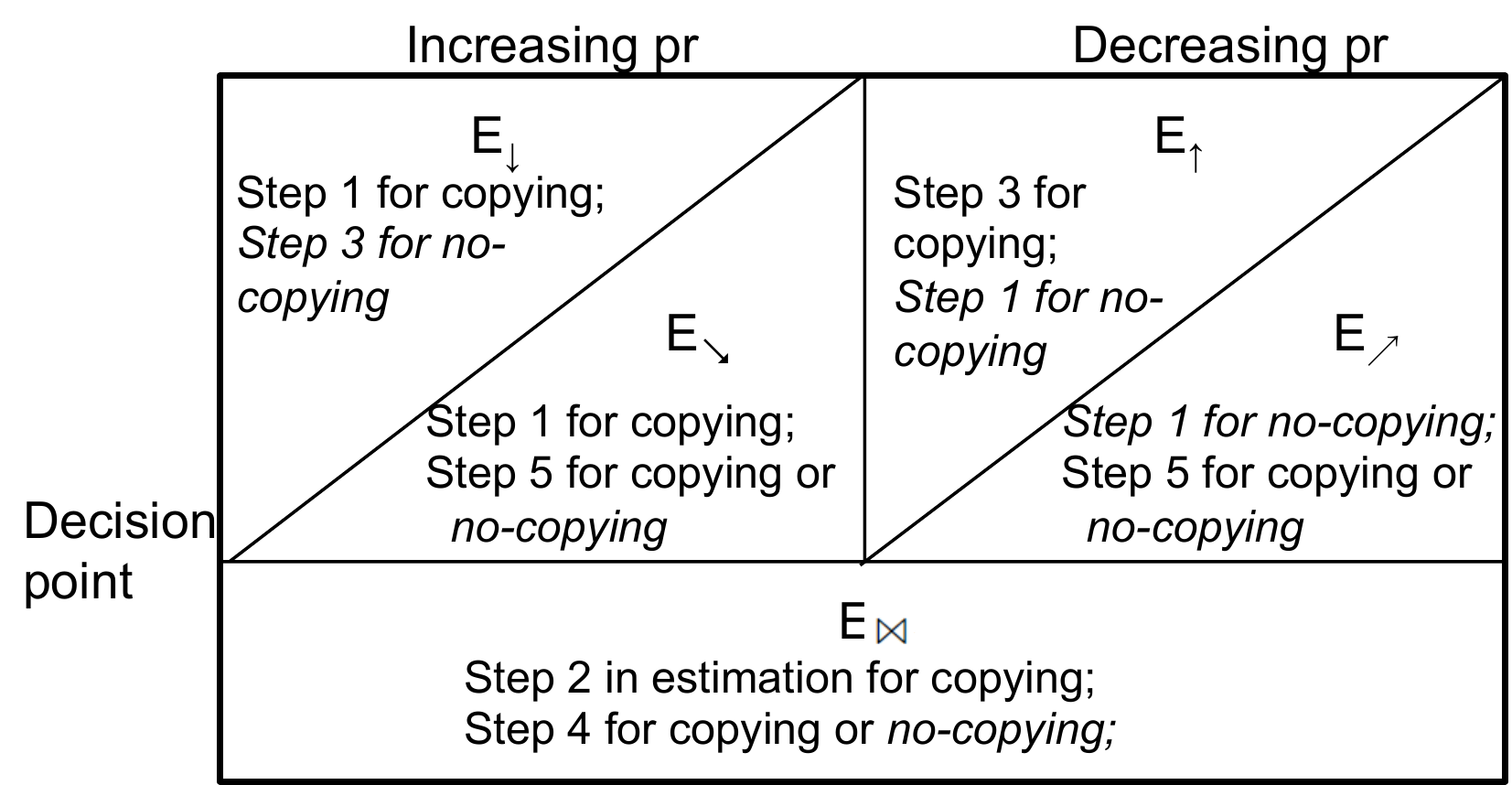}
{\small\caption{Entry categories. The steps for no-copying
pairs are in italic font.\label{fig:incremental}}}
\end{figure}

\subsection{Source pairs with copying}
\label{sec:onepair}
We first explain our strategy for source pairs where we concluded
with copying in the previous round. Recall that for a pair of sources, 
we may make our decision before reaching the end of the index;
we call the last entry we considered the {\em decision point}. 
Accordingly, we can categorize the shared entries between this pair of sources
into five categories (see Fig.\ref{fig:incremental}): 
(1) $\bar E_\downarrow$: big-change entries whose contribution 
scores decrease (the probabilities of the entries increase) before the
decision point; (2) $\bar E_\searrow$: small-change entries
whose scores decrease before the decision point; 
(3) $\bar E_\uparrow$: big-change entries whose scores increase 
before the decision point; (4) $\bar E_\nearrow$: small-change entries
whose scores increase before the decision point; 
and (5) $\bar E_\Join$: shared entries after the decision point. 

Among them, entries in $\bar E_\downarrow$ and $\bar E_\searrow$
would decrease the scores and may even change our decision.
The high-level idea for our algorithm is to first consider the decreases, and 
then compensate for score loss with the increases from the other categories until
the scores are once again above the threshold $\theta_{cp}$. 
In the latter process, we consider big increases
(from $\bar E_\uparrow$) first, and small ones (from $\bar E_\nearrow$) last. 
We next describe how we update $\hat C_\to$ (resp. $\hat C_\leftarrow$).

{\em Preparation step:} As a preparation, in the round when we 
conduct copy-detection from scratch, we maintain for each 
pair of sources the number of shared values before the decision point
and that after the decision point (the latter can be denoted by 
$|\bar E_\Join|$). We then compute the final score of this round, 
denoted by $\hat C_\to$ (resp. $\hat C_\leftarrow$),
as $\hat C_\to = C_\to^{min} - |\bar E_\Join| \cdot \ln(1-s)$.
Note that the computation of $C_\to^{min}$ assumes that there 
is no value shared after the decision point and applies penalty
$\ln(1-s)$ to each such shared value; 
the computation of $\hat C_\to$ removes this penalty but 
does not apply their real (positive) contribution; thus,
$C_\to^{min} < \hat C_\to < C_\to$.
We use $\hat C_\to$ and $\hat C_\leftarrow$
as the starting scores for the next round.

{\em Step 1 ($\bar E_\downarrow \cup \bar E_\searrow$):} 
Each entry $E \in \bar E_\downarrow$ may significantly reduce
$\hat C_\to$. We update $\hat C_\to$ by 
replacing the old score on $E$, computed by the old value
probability and source accuracy, with the new one computed by 
the new value probability and source accuracy.
Each entry $E \in \bar E_\searrow$ will reduce
$\hat C_\to$ slightly. Instead of updating
the change for each such entry, we use the maximum change,
denoted by $\Delta_\rho$, which we estimate from the entry 
with the largest score decrease below $\rho$.
We decrease $\hat C_\to$ by $\Delta_1=\Delta_\rho\cdot|\bar E_\searrow|$.
If after these changes $\max\{\hat C_\to, \hat C_\leftarrow\}\geq\theta_{cp}$ 
still holds, we can stop; otherwise, we conduct Step 2-5
and stop once $\max\{\hat C_\to, \hat C_\leftarrow\}\geq\theta_{cp}$.

{\em Step 2 ($\bar E_\Join$):} In case the new score
is below $\theta_{cp}$, we look for data entries that
can increase them back to above the threshold. 
Consider the shared entries after the decision point. 
Each of them should have a minimum contribution score, which can be
estimated on the last entry in the index.
We denote this score by $m$ and increase $\hat C_\to$ 
by $\Delta_2=m\cdot|\bar E_\Join|$.

{\em Step 3 ($\bar E_\uparrow$):} 
Each entry $E \in \bar E_\uparrow$ can significantly increase
$\hat C_\to$ and compensate for the loss. We update $\hat C_\to$ by 
replacing the old score on $E$ with the new one. 

{\em Step 4 ($\bar E_\Join$):} 
Each $E \in \bar E_\Join$ may also increase $\hat C_\to$ a lot.
We (1) increase $\hat C_\to$ by $C_\to(D_E)$, and
(2) subtract $m$ from $\hat C_\to$ and $\Delta_2$ to remove
our previous estimation on $E$.

{\em Step 5 ($\bar E_\searrow \cup \bar E_\nearrow$):}
Now the only entries not updated are those
with small changes before the decision point. 
For each $E \in \bar E_\searrow \cup \bar E_\nearrow$, 
we (1) update $\hat C_\to$ by 
replacing the old score on $E$ with the new one, and
(2) if $E \in \bar E_\searrow$, increase $\hat C_\to$ 
by $\Delta_\rho$ and subtract $\Delta_\rho$ from $\Delta_1$
to remove our previous estimation on $E$.

{\em Final step:} We remove the estimation, recording
$\hat C_\to+\Delta_1-\Delta_2$ as the precise $\hat C_\to$ 
(resp. $\hat C_\leftarrow$) for the starting point of the next round. 
We also update the new decision point if needed. 
If the condition $\max\{\hat C_\to, \hat C_\leftarrow\}\geq\theta_{cp}$
is not satisfied until the end,
we apply Bayesian analysis to decide if we need to change 
our decision to no-copying.

These steps can be combined into three passes of index
scanning. The first pass conducts Steps 1 and 2, the second
pass conducts Steps 3 and 4, and the third pass conducts Step 5.
Figure~\ref{fig:incremental} summarizes the algorithm.
We next illustrate the idea using an example. 

\begin{table}[t]
\vspace{-.15in}
\centering
{\small
\caption{Partial inverted index used in Round 3. Only source $S_0-S_4$
and their provided values are shown.}\label{tbl:incremental_index}
\begin{tabular}{c|p{1.1cm}|c|p{0.8cm}|p{0.35cm}|c}
\hline
Value & Providers & Pr & $\Delta$Score & Cat. & Score@$R_3$\\
\hline
TX.Houston  & $S_2, S_4$ & .04 $\to$ .03 & .17 & $\nearrow$ & 3.97 \\
FL.Miami & $S_2, S_3$ & .05 $\to$ .03 & .21 & $\nearrow$ & 3.83 \\
NJ.Atlantic  & $S_2, S_3, S_4$  & .07 $\to$ .03 & .39 & $\nearrow$ & 3.96 \\
{\em NY.Albany} & $S_0, S_1$ & $.07 \to .77$ & -$2.49$ & $\downarrow$ & $.52$ \\
NJ.Trenton  & $S_0, S_1$ & .9 $\to$ .95 & -.12 & $\searrow$ & 1.31 \\
{\em NY.NewYork}  & $S_2, S_3, S_4$ & $.84 \to .16$ & $1.69$ & $\uparrow$ & $3.17$ \\
AZ.Phoenix  & $S_0-S_4$ & .94 $\to$ .95 & -.01 & $\searrow$ & 1.45 \\
FL.Orlando  & $S_1, S_4$ & .9 $\to$ .92 & -.01 & $\searrow$ & .78 \\
TX.Austin & $S_0, S_1$ & .9 $\to$ .93 & -.01 & $\searrow$ & .51 \\
\hline
\end{tabular}}
\vspace{-.1in}
\end{table}
\begin{example} \label{eg:incre_copy}
Recall that for the motivating example there are five rounds before 
convergence (Table~\ref{tbl:iteration}).
Consider incremental detection at Round 3;
it considers value probabilities from Round 1 
as old ones and those from Round 2 as new ones. 
Table~\ref{tbl:incremental_index} shows the inverted index.
We set $\rho_V=1$, so there are 2 entries with big score changes (in italics),
corresponding to values for {\sf NY}.

First consider $(S_2,S_3)$. In Round 2 it terminates
at entry {\sf NY.NewYork}, having scores of $C_{\to/\leftarrow}^{min}=4.73$,
sharing 3 values before the decision point and 1 value after the point.
Thus, $\hat C_{\to/\leftarrow}=4.73+1.6=6.33$.
Among the shared entries before decision, 2 have small increases and 1 has 
big increase. Thus, the score is not decreased and
we can terminate for this pair without further examination.

Now consider $(S_0, S_1)$. In Round 2 it terminates 
at the last entry, having scores $C_\to=1.15, C_\leftarrow=1.66$,
and sharing 4 values before the decision point; recall that
$\theta_{cp}=2.08$ and $\theta_{ind}=1.39$, so we need to apply 
Eq.(\ref{eqn:ratio}), computing $P(S_0 \bot S_1|\Phi)=.32$
and deciding copying. Among the 4 shared values, {\sf NY.Albany} has
big score decrease and the other three have small decreases.
The contribution scores for $C_{\to/\leftarrow}$ from {\sf NY.Albany} were
$.44/1.53$ in Round 2 and are $.24/.09$ now.
The largest score difference for the other three items is .015, 
computed from {\sf NJ.Trenton}, which has the largest 
score decrease among small-change entries. Accordingly, we have
$\hat C_\to=1.15+(.24-.44)-.015*3=.9 < \theta_{cp}$,
and $\hat C_\leftarrow=1.66+(.09-1.53)-.015*3=.17 < \theta_{cp}$;
thus, we may change our decision.
Since $\bar E_\uparrow=\bar E_\Join=\emptyset$, we cannot compensate
for the loss of the score. We next reconsider the items with 
small changes and compute precise scores 
$\hat C_\to=.95 < \theta_{ind}, \hat C_\leftarrow=.20 < \theta_{ind}$. 
Therefore, we change our decision for this pair to no-copying. \rbox
\end{example}

\subsection{Source pairs with no-copying}
We handle source pairs with no-copying in a similar way.
For such pairs, entries in $\bar E_\uparrow$
and $\bar E_\nearrow$ would increase the scores and may
change our decision, while entries in $\bar E_\downarrow$
and $\bar E_\searrow$ would decrease the scores and compensate for
the score increase, so we change the order of considering them.
Also, Step 2 does not apply for no-copying pairs since we 
actually need to {\em reduce} the scores to compensate for its increase.
Again, the steps are summarized in Figure~\ref{fig:incremental}.
In addition, we compute $\hat C_{\to/\leftarrow}$ by Eq.(\ref{eqn:maxscore})
with two changes. First, we use the real number of different 
values obtained from bookkeeping rather than the estimated one. Second, 
in case the maximum score $M$ has a big change,
we update $\hat C_{\to/\leftarrow}$ upfront in each round.

\eat{
Finally, for an upper bound, if at its decision point
$M$ is larger than before and we cannot terminate, we need to 
conduct Step 4 and meanwhile re-consider the pruned pairs.
}
\begin{example}\label{eg:incre_indep}
Continue with Ex.\ref{eg:incre_copy} and now consider no-copying
pair $(S_0, S_2)$. In Round 2 it terminates at entry {\sf AZ.Phoenix},
having scores $C_\to^{max}=-4.75, C_\leftarrow^{max}=-4.3$, 
sharing 1 value before decision and 0 value after decision. 
The shared value is in category $\bar E_\searrow$, so Step 1
does not change the score and we can terminate with the same decision.\rbox
\end{example}

The final algorithm, {\sc Incremental}, updates scores for all source pairs
in three passes of index scanning. 
It requires more space for book-keeping across rounds,
but in practice it recomputes scores for much fewer entries.

\begin{proposition}
Let $r$ be the number of source pairs that share values,
and $e'$ be the maximum number of shared entries we process 
for each pair. {\sc Incremental} takes time 
$O(re')$ and space $O(|{\cal E}|+r+|{\cal S}|)$ 
for a single round. \rbox
\end{proposition}

\begin{example}\label{eg:incre}
In Rounds 3-5 for our example, 
{\sc Bound+} takes 102 computations for each round, while {\sc Incremental} reduces it to 54, 29 and 0 respectively. 
The total number of computations for {\sc Incremental}
is 73\% lower than that for {\sc Bound+}. \rbox
\end{example}



\section{Experimental Results}
\label{sec:experiment}
This section presents experimental results validating the efficiency
and effectiveness of the inverted index and algorithms proposed in this paper. 
We show that among the strategies we have proposed, the inverted
index can improve the efficiency by one to two orders of magnitude
and obtain exactly the same results;
pruning and incremental detection together can improve the efficiency
by nearly one order of magnitude and obtain very similar results;
and a careful sampling can improve the efficiency by orders of 
magnitude without sacrificing the quality of the results too much.

\begin{table}
\centering
\vspace{-.1in}
{\small
\caption{Overview of data sets.
\label{tbl:dataset}}
\begin{tabular}{|c|c|c|c|c|c|c|c|}
\hline
 & \#Srcs & \#Items & \#Dist-values & \#Index-entries \\
\hline
{\em Book-CS}   &   894   & 2,528 & 14,930 & 7,398 \\
{\em Stock-1day}   &   55   &  16,000  & 104,611 & 40,834          \\
{\em Book-full}   &   3,182   &  147,431  & 162,961 & 48,683          \\
{\em Stock-2wk}   &   55   &  160,000  & 915,118 & 405,537          \\
\hline
\end{tabular}}
\vspace{-.05in}
\end{table}
\begin{table*}
\centering
{\small
\caption{Copy-detection and truth-discovery quality of various
algorithms. Except fusion accuracy, all measures are computed
by comparing with results of {\sc Pairwise}. {\sc Sample2} obtains
the same results as {\sc Sample1} on {\em Stock} data. 
\label{tbl:correctness}}
\begin{tabular}{|c||c|c|c|c|c|c||c|c|c|c|c|c|}
\hline
\multirow{3}{*}{Method} & \multicolumn{6}{c||}{\em Book-CS} & \multicolumn{6}{c|}{\em Stock-1day}  \\
\cline{2-13}
& \multicolumn{3}{c|}{Copy detection} & \multicolumn{3}{c||}{Truth discovery} & \multicolumn{3}{c|}{Copy detection} & \multicolumn{3}{c|}{Truth discovery} \\
\cline{2-13}
& Prec & Rec & F-msr & Accu & Fusion diff & Accu var 
& Prec & Rec & F-msr & Accu & Fusion diff & Accu var \\
\hline
{\sc Pairwise} & - & - & - & .890 & - & - & - & - & - & .897 & - & -\\
{\sc Sample1} &.691 & .165& .264&.870 &.070 &.127 &.967& .945&.956&.896&.008&.001\\
{\sc Sample2}  &.886 & .696& .779 & .880 & .029& .089&.967& .945&.956&.896&.008&.001\\
\hline
{\sc Index}& 1 & 1& 1 & .890 & 0 & 0 & 1 &1&1&.897&0&0\\
 {\sc Hybrid} &.990& .980&.985 & .890& .015& .039&1 &.970 &.985&.897&.002&.001\\
  {\sc Incremental}&.985& .975&.980 &.890 & .015& .037&.993 &.947 &.969&.897&.003&.001\\
 {\sc ScaleSample}&.930 & .841& .882 & .890 & .029& .055& .970& .927 &.948&.897&.008&.001\\
\hline
\end{tabular}}
\end{table*}

\begin{table*}
\centering
\vspace{-.1in}
{\small
\caption{Execution time and the time improvement
compared with the previous method ({\sc Sample1, Sample2, Index}
comparing with {\sc Pairwise}; others comparing with
the method in the above row). {\sc Sample2} obtains
the same results as {\sc Sample1} on {\em Stock} data. 
\label{tbl:efficiency}}
\begin{tabular}{|c||c|c||c|c||c|c||c|c|}
\hline
\multirow{2}{*}{Method} & \multicolumn{2}{c||}{\em Book-CS} & \multicolumn{2}{c||}{\em Stock-1day} & \multicolumn{2}{c||}{\em Book-full} & \multicolumn{2}{c|}{\em Stock-2wk}  \\
\cline{2-9}
& Time (s) & Improvement & Time (s) & Improvement & Time (s) & Improvement & Time (s) & Improvement \\
\hline
{\sc Pairwise} & 321 & - & 306 & - & 11536 & - & 3408 & - \\
{\sc Sample1} & 3.2 & 99\% & 16.2 & 95\% & 278 & 98\% & 55 & 98\% \\ 
{\sc Sample2} & 32 & 90\% &16.2 &95\% & 684 & 94\% & 55 & 98\%  \\
\hline
{\sc Index} & 1.6 & 99.5\% & 25.0 & 92\% & 47.7 & 99.6\% & 573 & 83\% \\
{\sc Hybrid} & 1.2 & 24\% & 15.8 & 37\% & 47.2 & 2\% & 443 & 23\% \\
{\sc Incremental} & 0.4 & 65\% & 6.9 & 56\% & 7.9 & 83\% & 127 & 72\% \\
{\sc ScaleSample} & 0.3 & 25\% & 0.7 & 90\% & 3.8 & 52\% & 1.4 & 99\% \\
\hline
{\bf Total Improvement} & & {\bf 99.91\%} & & {\bf 99.8\%} & & {\bf 99.97\%} & & {\bf 99.96\%} \\
\hline
\end{tabular}}
\vspace{-.1in}
\end{table*}

\subsection{Experiment settings}
\label{sec:exper_settings}
\noindent
{\bf Data:} We experimented on four data sets\footnote{\small
The data are at {\em http://lunadong.com/fusionDataSets.htm}.};
Table~\ref{tbl:dataset} provides an overview. 
Two data sets were crawled from an online bookstore aggregator {\em AbeBooks.com}:
{\em Book-CS} contains 894 sources (\ie, book stores), 1265 CS books, 
and 2528 data items including the title and author list of each book 
(there are missing values for some books); on average 5.9 conflicting values
are provided for each data item.
{\em Book-full} contains 3182 sources, 81,352 books of all categories,
and 147,431 data items; on average 1.1 conflicting values are provided
for each data item. A gold standard for {\em Book-CS} contains
    author lists verified from book title pages for 100 randomly 
    selected books. 

The other two data sets were crawled from 55 Deep Web sources
on 16 attributes of 1000 stocks.
{\em Stock-1day} includes the data on 7/7/2011
and {\em Stock-2wk} includes the data from 7/1/2011 to 7/14/2011.
The former contains $16*1000=16,000$ data items and on average 
6.5 conflicting values are provided for each data item; 
the latter contains $16,000*10=160,000$ data items and on average
5.7 conflicting values are provided for each item. 
A gold standard for {\em Stock-1day} contains 
the voting results on the 100 NASDAQ symbols
    and 100 other randomly selected symbols 
    from 5 popular financial websites: {\em NASDAQ, Yahoo! Finance,
    Google Finance, MSN Money,} and {\em Bloomberg}.

The four data sets have very different features. 
{\em Book-full} and {\em Stock-2wk} contain a large number of data items.
{\em Book-CS} and {\em Book-full} contain a large number of data sources;
however, some sources contain only a few data items 
(\eg, 85\% sources in {\em Book-CS} each covers at most 1\% books).
{\em Stock-1day} and {\em Stock-2wk} contain
much fewer sources, but each source has a much higher coverage
(\eg, 80\% sources each covers over half of the data items). 

\smallskip
\noindent
{\bf Implementation:} We implemented various methods for copy detection
and describe them as follows.
\begin{itemize}
  \item {\sc Pairwise} examines each pair of sources as described
    at the beginning of Section~\ref{sec:opportunity}~\cite{DBS09a}.
  \item {\sc Sample1} randomly samples 1\% of data items on {\em Stock-2wk}
    and 10\% on the other data sets, then applies {\sc Pairwise}
    on the sampled data.
  \item {\sc Sample2} is different from {\sc Sample1} on
    the two {\em Book} data sets. 
    It considers each data set as a table where
    each row represents a source and each column represents a
    data item. It randomly samples data items (columns) until the
    number of non-empty cells reaches 65\% on {\em Book-CS} 
    and 24\% on {\em Book-full} (we explain the need for such sampling
    rates shortly). 
  \item {\sc Index} implements algorithm {\sc Index} 
    (Section~\ref{sec:index}).
  \item {\sc Bound} and {\sc Bound+} each applies the corresponding
    algorithm (Section~\ref{sec:singleround}) for each round.
  \item {\sc Hybrid} applies {\sc Index} for a pair of sources that
    share at most 16 data items\footnote{\small We observe empirically
    that when two sources share fewer than 16 data items, {\sc Index}
    conducts fewer computations than {\sc Bound+} on average.} 
    and applies {\sc Bound+} for other
    pairs in each round (end of Section~\ref{sec:singleround})
  \item {\sc Incremental} applies {\sc Hybrid} in the first two
    rounds and Algorithm {\sc Incremental} 
    (Section~\ref{sec:incremental}) in later rounds. \footnote{\small Empirically we found that copy-detection and truth-finding results vary a lot in the first two rounds in general, so applying INCREMENTAL in the second round would not save much.}
    It sets $\rho$ to $.2$ for source accuracy and to $1.0$ for 
    value probability according to observations of the largest gaps
    on differences of changes.
  \item {\sc ScaleSample} applies {\sc Incremental}
    on a sampled data set, where we sample 1\% of data items on {\em Stock-2wk}
    and 10\% on the other data sets, and guarantee sampling at least
    $N=4$ data items from each source.
  \item {\sc FaginInput} generates the input to Fagin's NRA
   algorithm as described at the end of Section~\ref{sec:opportunity}.
\end{itemize}

In addition, we used the truth-finding algorithm 
in~\cite{DBS09a}, which considers both copying and source accuracy.
We plugged in the aforementioned copy-detection algorithms.

We implemented the algorithms in Java on a Windows machine with Intel Core 
i5 processor (3.2GHz, 4MB cache, 4.8 GT/s QPI, 8GB memory).

\smallskip
\noindent
{\bf Measures:} We measure three aspects of different methods.

\noindent
{\em Efficiency:} We measure efficiency by 
    (1) the number of computations in copy detection (as described
    in the examples in Sections~\ref{sec:index}-\ref{sec:incremental}), 
    and (2) the execution time.

\smallskip
\noindent
{\em Copy-detection correctness:} We examined how the
    various methods for improving scalability may hurt
    the results of copy detection; thus, we compared their results
    with those of {\sc Pairwise}. {\em Precision} measures
    among the output copying pairs, what fraction is also
    output by {\sc Pairwise}; {\em Recall} measures among the
    output copying pairs by {\sc Pairwise}, what fraction
    is output by the specific method; {\em F-measure} is computed by
    $2\cdot precision\cdot recall \over precision+recall$. 

\smallskip
\noindent
{\em Truth-finding correctness:} We also examined how the
    copy-detection results may affect truth finding.
    We report three measures: (1) {\em Fusion accuracy} measures
    the fraction of correct truth-finding results among all 
    data items in the gold standard; (2) {\em Fusion difference} 
    measures the fraction of truth-finding results different 
    from those when applying {\sc Pairwise}; and 
    (3) {\em Accuracy variance} measures the average difference
    of the source accuracies we compute
    when applying {\sc Pairwise} and the specific
    copy-detection method.

We report efficiency on all data sets and other results only
on the two small data sets {\em Book-CS} and {\em Stock-1day}. 

\eat{
\begin{table}
\centering
\vspace{-.1in}
{\small
\caption{Execution time (in ms) of each component of {\sc ScaleSample}.
\label{tbl:splittime}}
\begin{tabular}{|c|c|c|c|c|}
\hline
Component & {\em Book-CS} & {\em Stock-1day} & {\em Book-full} & {\em Stock-2wk} \\
\hline
Sampling & 11 & 47 & 1624 & 452 \\
Indexing & 108 & 217 & 925 & 339 \\
Detection & 200 & 407 & 1285 & 637 \\
\hline
\end{tabular}}
\vspace{-.1in}
\end{table}
}
\eat{
\begin{figure*}[t]
\begin{minipage}[b]{0.24\textwidth}
\center
\includegraphics[width=1.7in]{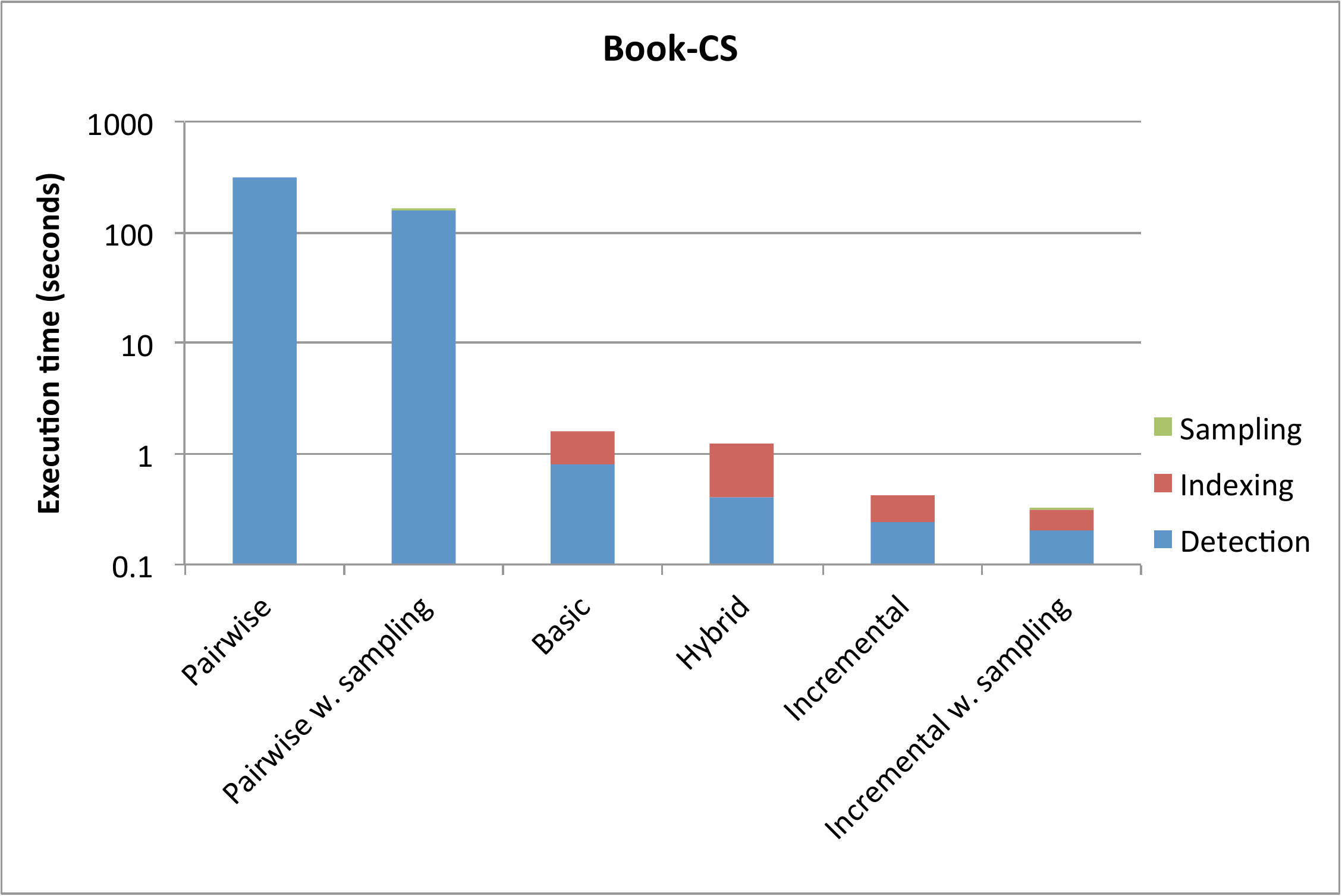}
\end{minipage}
\begin{minipage}[b]{0.24\textwidth}
\center
\includegraphics[width=1.7in]{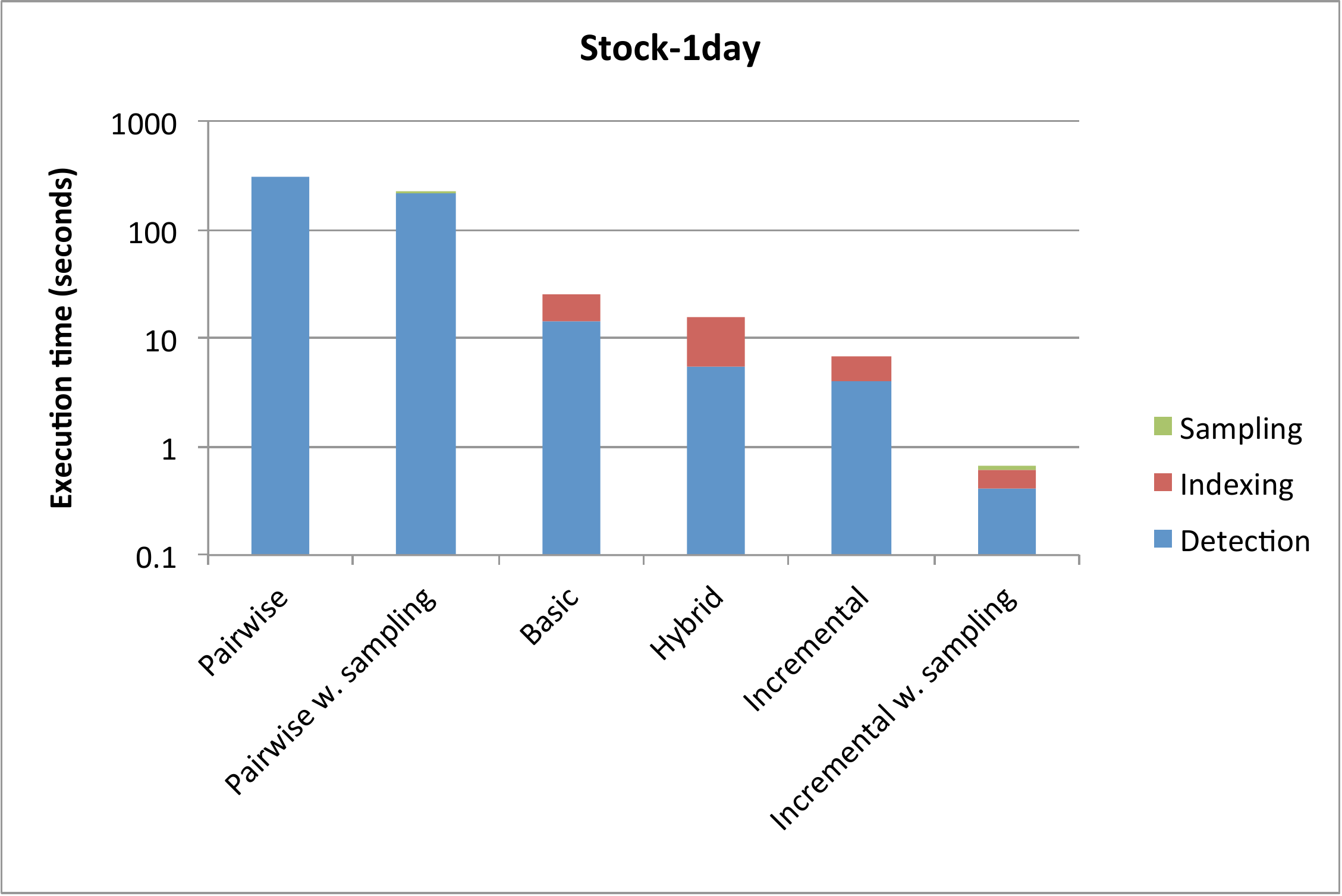}
\end{minipage}
\begin{minipage}[b]{0.24\textwidth}
\center
\includegraphics[width=1.7in]{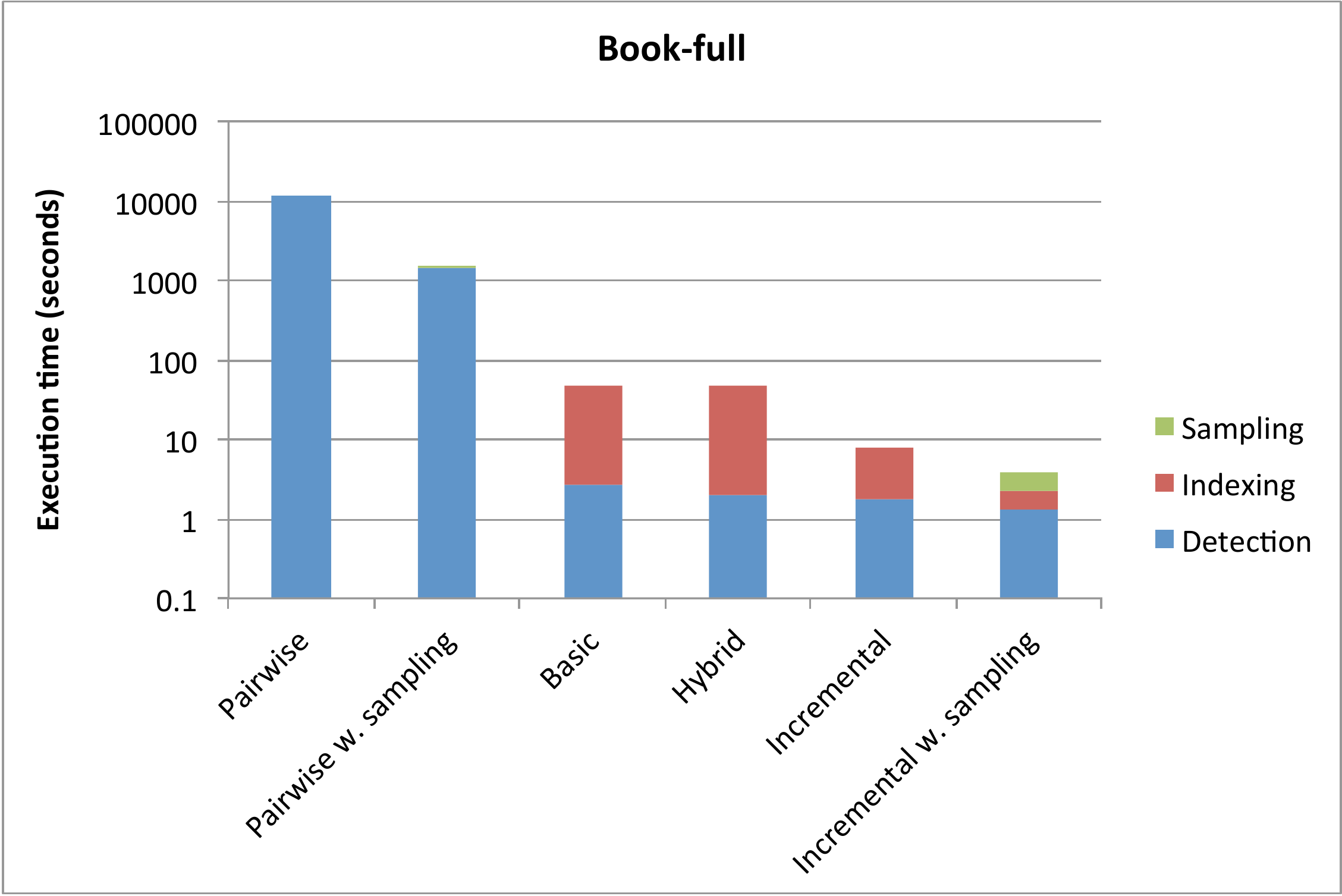}
\end{minipage}
\begin{minipage}[b]{0.24\textwidth}
\center
\includegraphics[width=1.7in]{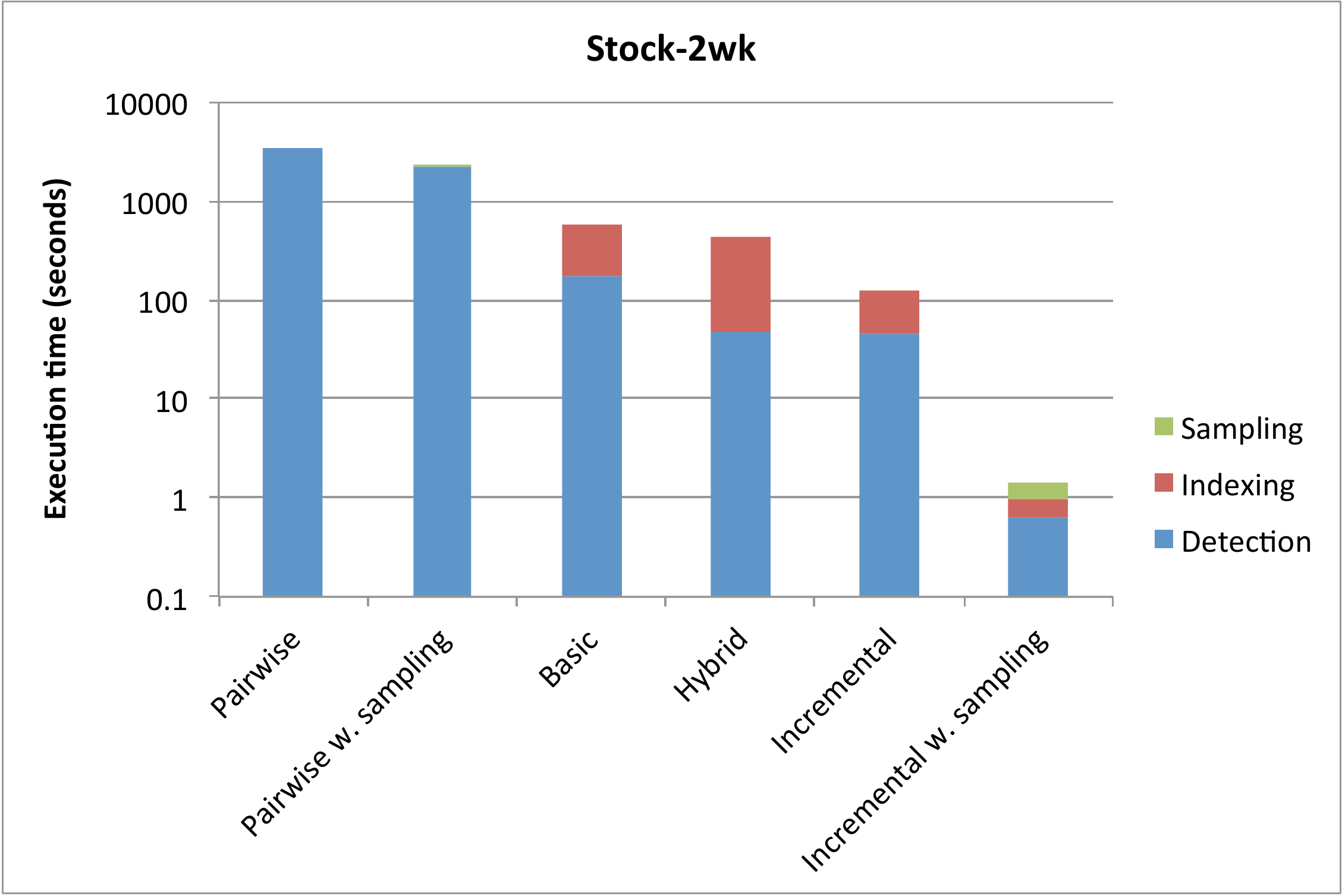}
\end{minipage}
\vspace{-.1in} 
\caption{\small Execution time of each component of various algorithms.
\label{fig:splittime}}
\end{figure*}}
\subsection{Performance overview}
\label{sec:exper_overview}
We first compare the various methods on each data set. 
Table~\ref{tbl:correctness} reports copy-detection and
truth-finding correctness, and Table~\ref{tbl:efficiency} reports
execution time. 

First, naive sampling ({\sc Sample1} and {\sc Sample2}) 
did improve the efficiency a lot, but not as much as 
{\sc Incremental} and {\sc ScaleSample}.
Indeed, on the {\em Stock} data sets they are one order of magnitude slower than 
{\sc ScaleSample} and on the {\em Book} data sets they are 
even slower than {\sc Index}. 
In addition, {\sc Sample1} obtains very low F-measure on copy
detection for {\em Book-CS}, where a lot of
data sources provide only a few books, so a random sampling
can lead to inaccurate decisions.

Second, our proposed methods for improving scalability work very well.
Without sampling, {\sc Incremental} finished in about 2 minutes for 
{\em Stock-2wk} and seconds for other data sets.
In particular, the use of the inverted index in itself ({\sc Index}) 
on average reduced execution time by 94\% and obtains exactly
the same results for copy detection and truth discovery as {\sc Pairwise}.
It works especially well for the two {\em Book} data sets 
(improving by two orders of magnitude) because a lot of source
pairs (95.6\% on average) do not share any data item and 
need not to be considered at all. Also, we observe from 
Table~\ref{tbl:dataset} that on average only 42\% values are
provided by multiple sources and so are indexed.
Pruning ({\sc Hybrid}) on average reduced execution time further
by 21\% and changed copy-detection and truth-discovery results 
very slightly. Incremental detection ({\sc Incremental}) on average
reduced execution time further by 69\% and also changed the results
very slightly. The two enhancements together reduced execution time 
by 77\% on average and sacrificed precision and recall of copy detection 
by at most 5\%; they also changed results of truth-discovery 
very slightly, by up to 1.5\%. We observed from our experiments
that indexing costs 57\% of execution time in {\sc Incremental},
but it spent only .9\% execution time of {\sc Pairwise} and
significantly improves scalability, so is worthwhile.

Third, sampling helps with a small sacrifice on effectiveness: 
{\sc ScaleSample} finished within a few seconds for all data sets 
with reasonable F-measure for
copy detection and very similar results for truth discovery.
On the {\em Stock} data sets, the improvement corresponds to the sampling
rate: 90\% for {\em Stock-1day} (sampling rate .1) and 
99\% for {\em Stock-2wk} (sampling rate .01); in addition, 
the F-measure and fusion results are very similar to 
{\sc Incremental}, which does not do sampling.
On the {\em Book} data sets, the efficiency was improved 
but not as much (by 25\% and 52\% respectively), and the F-measure 
of copy detection drops. Recall that in these two data sets
there are a lot of low-coverage sources,
making sampling much harder. Indeed, 
we ended up sampling 49\% data items for {\em Book-CS}
and 19\% items for {\em Book-full}. 
However, we obtain much higher F-measure than 
{\sc Sample1} and {\sc Sample2}, showing effectiveness of 
sampling at least $N=4$ data items. Last, we note that sampling
in itself has a very small overhead for small data sets
(5\% of execution time on average) but a larger overhead for large data
sets (37\% on average); this is because checking whether each source
covers $N$ sampled data items takes longer time for large
data sets. 

Finally, Table~\ref{tbl:fagin} shows the execution time ratio
of our methods versus {\sc FaginInput}. {\sc FaginInput} has two
drawbacks. First, it has to compute the contribution scores from
each shared value for \emph{ each source pair}; thus, {\sc Hybrid} is 18\%
faster than {\sc FaginInput} on average for a single round. 
Second, it is not clear how to generate the input lists 
\emph{incrementally in later rounds}; thus, {\sc Incremental} is 75\% faster 
than {\sc FaginInput} on average for all rounds. 

\eat{
\begin{table}
\centering
{\small
\caption{Split time cost of Hybrid (ms)
\label{tbl:splittime_hybrid}}
\begin{tabular}{|c|c|c|c|c|}
\hline
           &Book-CS & Stock-1day& Book-full& Stock-2wk \\
\hline
Detection  &   411   & 5600 & 2015 & 47964 \\
Indexing   &   825   &  10200  & 45259 & 394647         \\
\hline
\end{tabular}}
\vspace{-.15in}
\end{table}
}

\begin{figure*}[t]
\begin{minipage}[t]{0.5\textwidth}
\includegraphics[width=1.7in, height=1.25in]{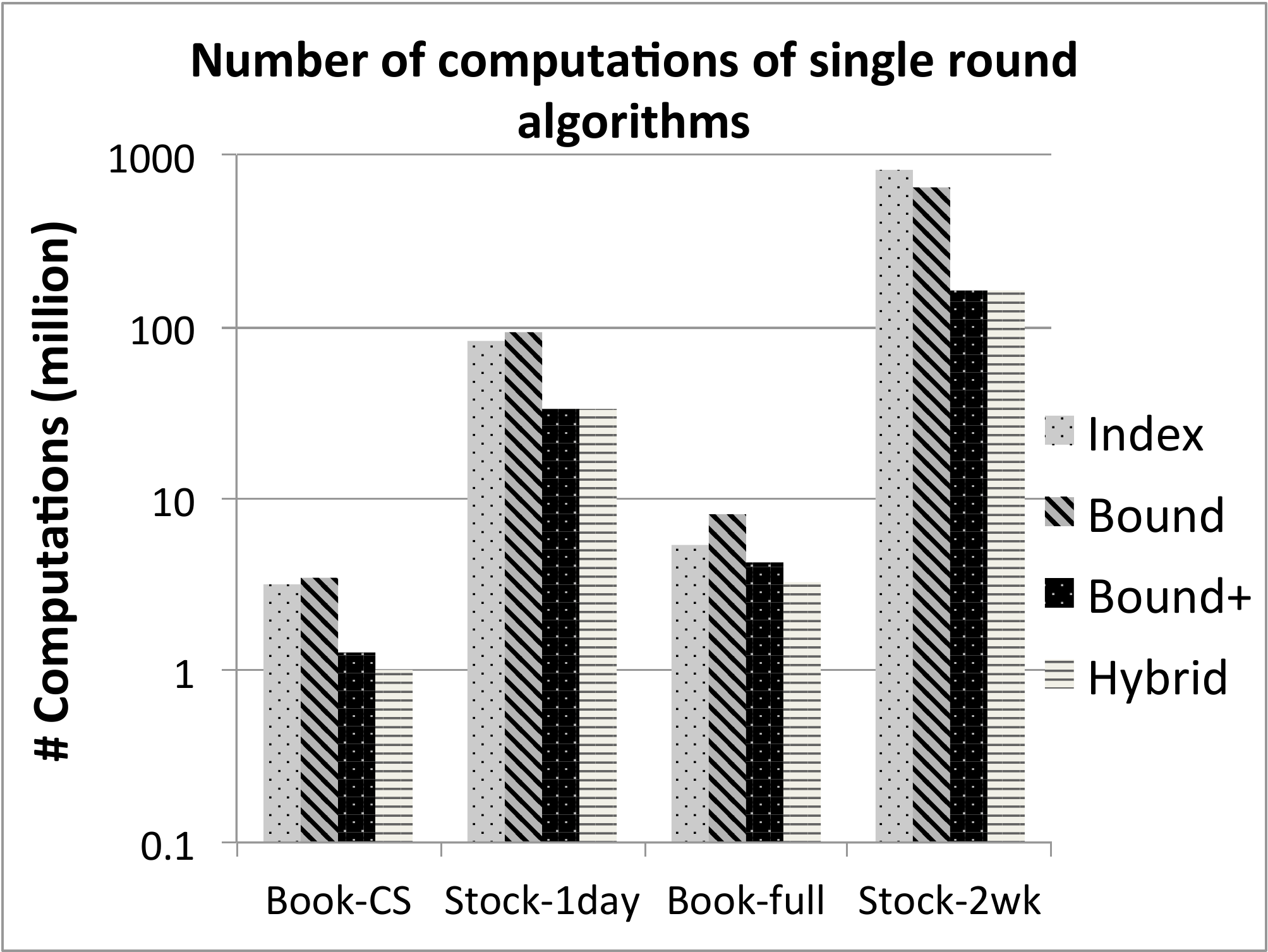}
\includegraphics[width=1.7in, height=1.25in]{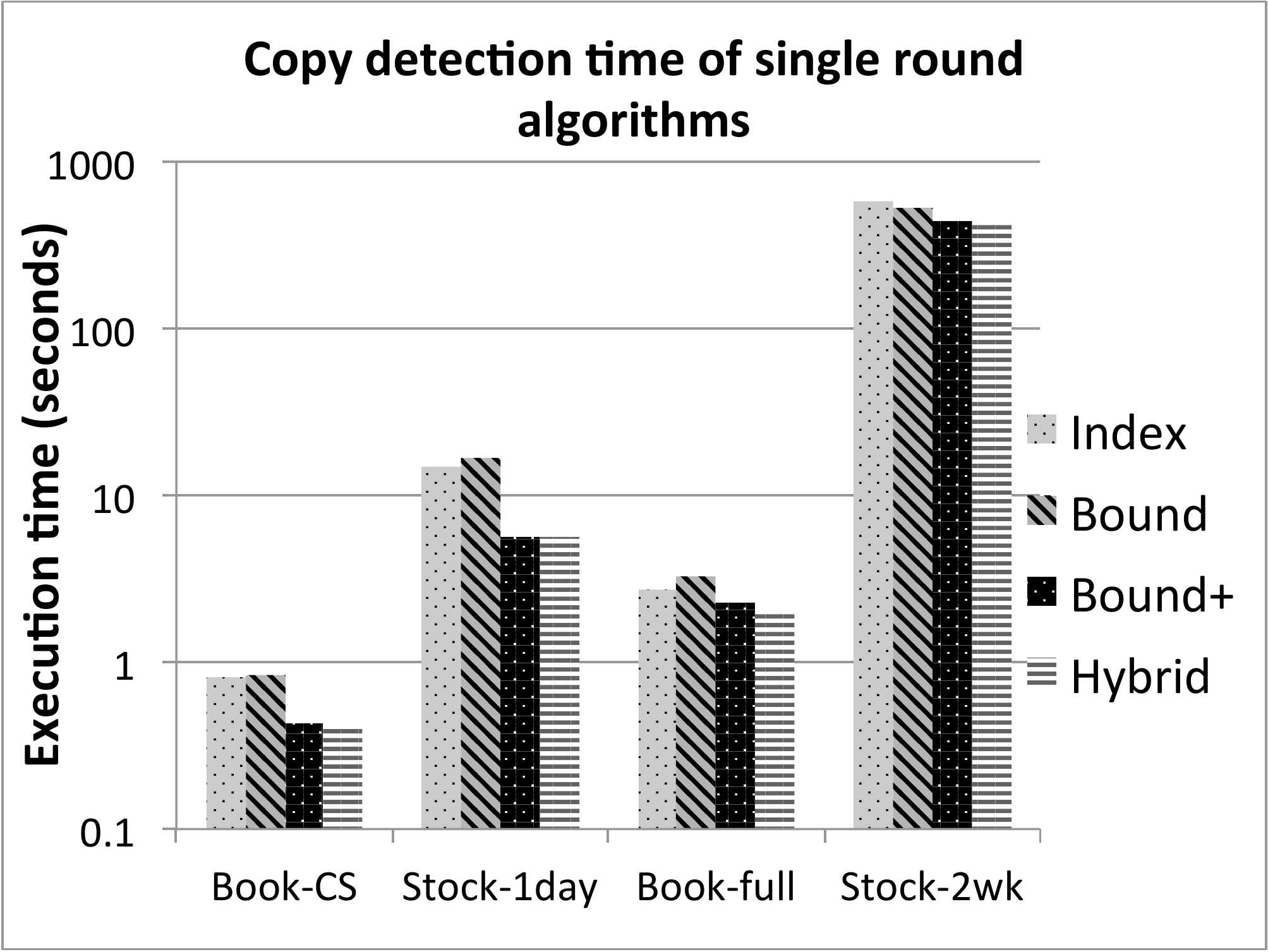}
{\small\caption{Single-round algorithms.\label{fig:singleround}}}
\end{minipage}
\hfill
\begin{minipage}[t]{0.5\textwidth}
\includegraphics[width=1.7in, height=1.25in]{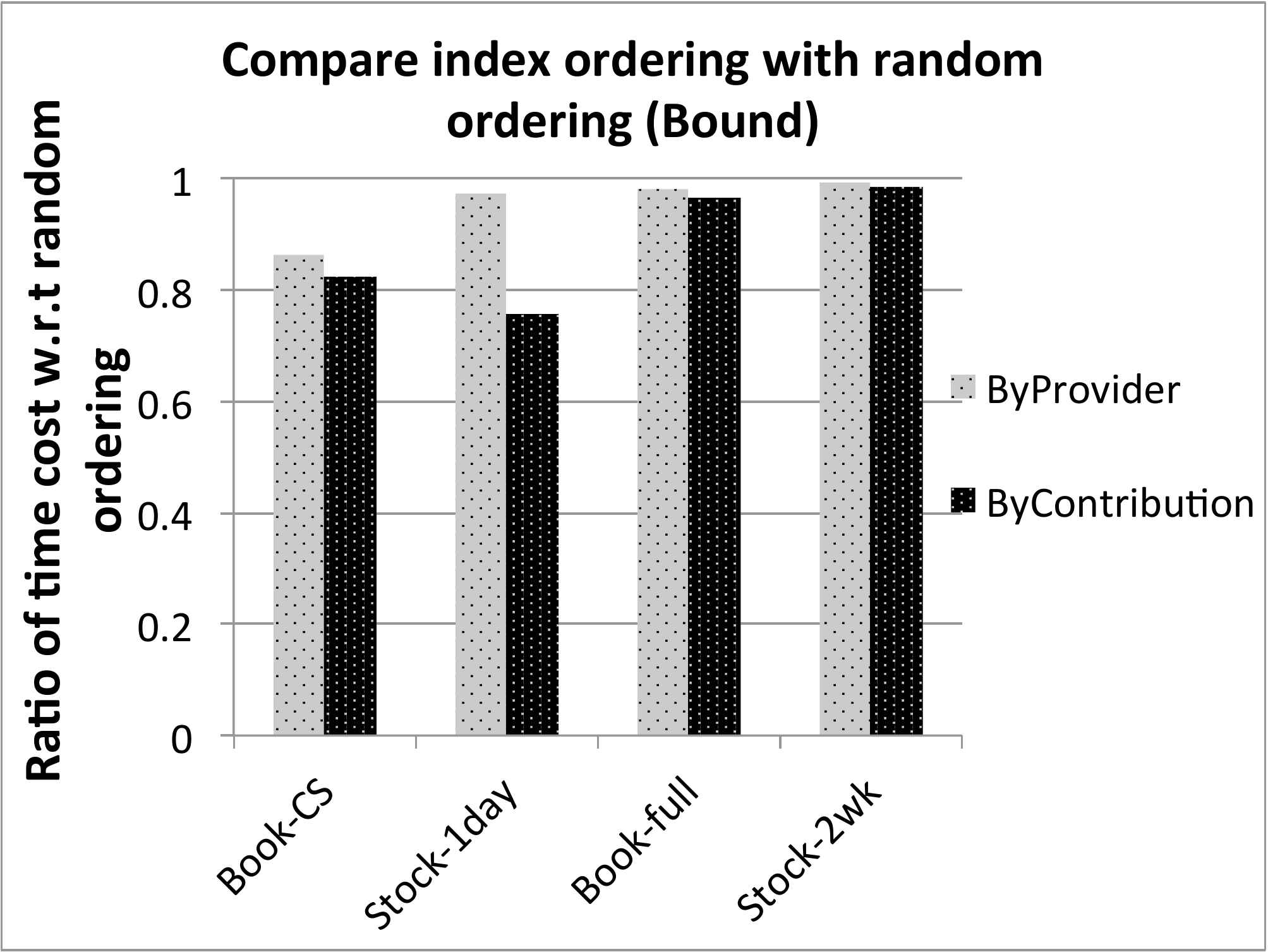}
\includegraphics[width=1.7in, height=1.25in]{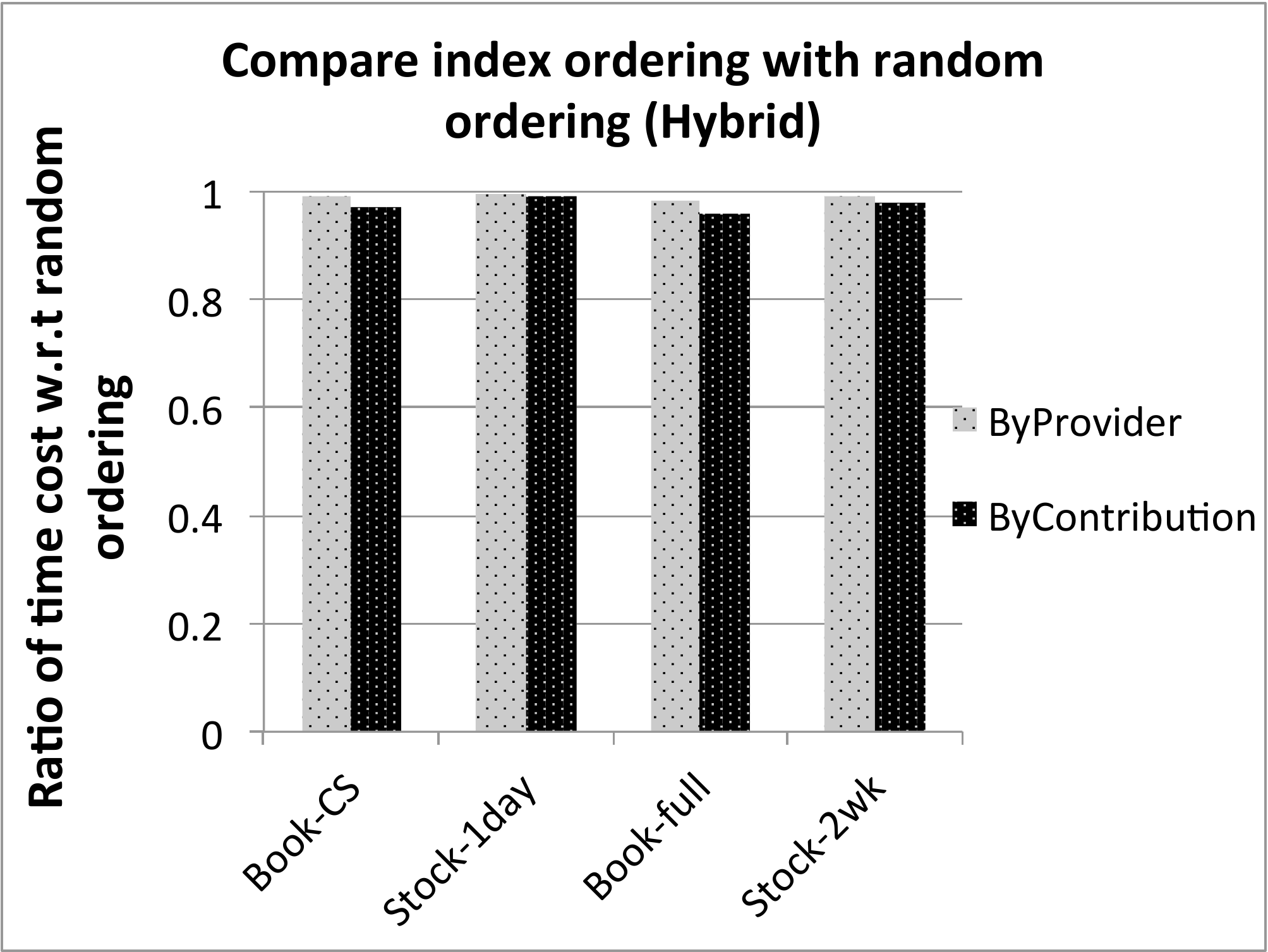}
{\small\caption{Different index ordering.\label{fig:index}}}
\end{minipage}
\end{figure*}
\subsection{Single-round algorithms}
\label{sec:singleRoundExp}
We next examine single-round algorithms in more detail.
We first compare {\sc Index}, {\sc Bound}, {\sc Bound+},
and {\sc Hybrid} on their numbers of computations (for all
rounds together) and copy-detection time
(see Figure~\ref{fig:singleround}). We have three observations.
First, for three out of 
four data sets {\sc Bound} conducts more computations 
and finished in longer time than {\sc Index}. Although 
it reduces the number of data items for consideration,
it introduces a big overhead for computing the minimum and
maximum scores. Second, {\sc Bound+} speeds up copy detection
significantly: on average it reduces the number of computations by
55\% and saves copy-detection time by 37\% over {\sc Bound}.
Third, {\sc Hybrid} further saves 20.3\%, 22.9\% computations 
and 4.6\%, 11.6\% copy-detection time on {\em Book-CS} and 
{\em Book-full} respectively. It does not make a difference
on the two {\em Stock} data sets, because there each pair of 
sources share a lot of data items.

We then examined various orders of processing entries in the inverted index:
{\sc Random} processes the entries 
randomly; {\sc ByProvider} processes the entries in increasing order
of the number of providers (\ie, sources); and {\sc ByContribution} processes the
entries in decreasing order of contribution (proposed in this paper). 
Figure~\ref{fig:index} shows the execution
time of each of the latter two compared with random ordering
for {\sc Bound} and {\sc Hybrid}. 
We observed that {\sc ByContribution} is the fastest among
the three ordering schemes. When we apply {\sc Bound},
it improves over {\sc Random} by 
12\% on average and by 24\% for {\em Stock-1day}; it improves
over {\sc ByProvider} by 7\% on average and by 22\% for 
{\em Stock-1day}. When we apply {\sc Hybrid}, which skips many
computations by setting up a timer, the benefit of 
{\sc ByContribution} is less evident but it is still the fastest.
We also note that although {\sc ByProvider} is better than 
{\sc Random}, it may process some true but not widely provided
values towards the beginning and so can incur more
computation than {\sc ByContribution}.

\begin{table}
\vspace{-.1in}
\centering
{\small
\caption{Execution time ratio of {\sc Incremental} vs {\sc Hybrid},
percentage of pairs terminated at each pass of incremental detection
\label{tbl:incremental_details}}
\begin{tabular}{|c|c|c|c|c|}
\hline
& {\em Book-CS} & {\em Stock-1day} & {\em Book-full} & {\em Stock-2wk}\\
\hline
Round 3 & 14.0\% & 6.9\% & 3.1\% & 7.3\% \\
Round 4 & 12.2\% & 6.8\% & 3.3\% & 4.7\% \\
Round 5 & 10.2\% & 6.1\% & 3.4\% & 4.4\% \\
Round 6 & 9.6\%  & 6.4\% & 3.3\% & 4.9\% \\
Round 7 & 10.2\% & - & 3.7\% & - \\
Round 8 & 9.6\%  & - & 3.1\% & - \\
Round 9 & - & - & 3.0\% & - \\
\hline
\hline
Pass 1& 99\% & 98\% & 86\% & 99\%\\
Pass 2& 0 & 1\% & 4\% & 0\\
Pass 3& 1\% & 1\% & 10\% & 1\%\\
\hline
\end{tabular}}
\vspace{-.15in}
\end{table}
\begin{table}
{\small
\caption{Comparing different sampling methods.
\label{tbl:sampling}}
\begin{tabular}{|c|c|c|c|c|c|c|}
\hline
&\multicolumn{3}{|c|}{\em Book-CS}  & \multicolumn{3}{|c|}{\em Stock-1day} \\
\hline
Method & Prec & Rec & F-msr & Prec & Rec & F-msr\\
\hline
{\sc ScaleSample} & .92 & .84& .88 & .98 &.94&.96\\
{\sc ByItem}&.85 & .56& .67 &.98& .94&.96\\
{\sc ByCell} &.89 & .70& .78 & .98& .94& .96\\
\hline
\end{tabular}}
\vspace{.1in}
{\small
\caption{Execution-time ratio w.r.t. {\sc FaginInput}. 
\label{tbl:fagin}}
\begin{tabular}{|c|c|c|c|c|}
\hline
           &{\em Book-CS} & {\em Stock-1day} & {\em Book-full}&{\em Stock-2wk} \\
\hline
{\sc Hybrid} & .87 & .76 & .99 & .67 \\
{\sc Incremental} & .30 & .27 & .22 & .19 \\
\hline
\end{tabular}}
\end{table}
\subsection{Incremental algorithms}
\label{sec:exper_incremental}
To understand how incremental detection improves efficiency, 
we show in Table~\ref{tbl:incremental_details} the execution time
ratio of {\sc Incremental} versus {\sc Hybrid} round by round. 
Indeed, incremental detection saves execution
time significantly: on average it improves over {\sc Hybrid} 
by 97\% for indexing, 52\% for copy detection, 
and 93.5\% in total.
We also show in Table~\ref{tbl:incremental_details} how many pairs 
terminate at each of the three passes 
We observe that in the first pass 86\% pairs terminate for {\em Book-full}
and over 98\% pairs terminate for other data sets.
This verifies our intuition and explains why {\sc Incremental} 
can save computation significantly.

\eat{
\begin{table}
\centering
{\small
\caption{Comparing different sampling methods.
\label{tbl:sampling}}
\begin{tabular}{|c||c|c|c||c|c|c|}
\hline
\multirow{3}{*}{Method} & \multicolumn{3}{c||}{\em Book-CS} & \multicolumn{3}{c|}{\em Stock-1day}  \\
\cline{2-7}
& Prec & Rec & F-msr & Prec & Rec & F-msr \\
\hline
{\sc ScaleSample} & .92 & .84& .88 & .98 &.94&.96\\
{\sc ByItem}&.85 & .56& .67 &.98& .94&.96\\
{\sc ByCell} &.89 & .70& .78 & .98& .94& .96\\
\hline
\end{tabular}}
\vspace{-.15in}
\end{table}}
\subsection{Sampling}
\label{sec:exper_sampling}
Finally, we compare our sampling strategy, called {\sc ScaleSample},
with sampling rate 10\%,
with two naive sampling strategies as described in 
{\sc Sample1} and {\sc Sample2},
which we call {\sc ByItem} and {\sc ByCell} respectively;
here we apply {\sc Incremental} on all samples.
To ensure a fair comparison, the sampling rate 
for {\sc ByItem} is decided by the percentage of sampled data items 
in {\sc ScaleSample}, and the sampling rate for {\sc ByCell}
(and {\sc Sample2}) is decided
by the percentage of sampled cells in {\sc ScaleSample}. For example,
{\sc ScaleSample} sampled 49\% data items and 65\% cells on 
{\em Book-CS}, so we applied a sampling rate of 49\% for {\sc ByItem}
and 65\% for {\sc ByCell};
{\sc ScaleSample} sampled 10\% data items and 10\% cells on 
{\em Stock-1day}, so {\sc ByItem} and {\sc ByCell} applied the same 
sampling rate (10\%). 
Table~\ref{tbl:sampling} shows
the quality of copy-detection results compared to applying {\sc Index}. 
The three sampling methods
obtain the same results on {\em Stock-1day} since the sources all
have a high coverage in that data set; {\sc ScaleSample} obtains
the best results on {\em Book-CS} even though it selects 
the same number of data items
as {\sc ByItem} and the same number of cells as {\sc ByCell},
since it guarantees that we select at least $N=4$ data items 
from each source when possible.

\eat{
We also examined the effect of sampling rate and the value of $N$;
Figure~\ref{fig:sampling} shows the results of {\sc ScaleSample}.
First, the percentage of sampled data items is exactly the
same as the sampling rate for {\em Stock-1day}, but much
higher than the sampling rate for {\em Book-CS}; in other words,
$N$ plays a big role when there are a lot of low-coverage sources.
Second, as expected, the higher the sampling rate, the higher the
F-measure of copy detection. \eat{On {\em Stock-1day}, we observe 
a big jump when the sampling rate grows from .05 to .1
and then the increase slows down. In contrast, on {\em Book-CS} 
we obtain similar F-measures when the sampling rate is .05 and is .1,
since {\sc ScaleSample} is controlled more by $N$ when 
sampling rate is low on this data set. }Third, on {\em Book-CS}
when $N$ increases, the F-measure of copy detection increases as well.
We observe a big jump when $N$ grows from 2 to 4 and then
the increase slows down. How to set the sampling rate and the value of $N$
is a trade-off between efficiency and effectiveness. We observe
that a sampling rate of .1 and $N=4$ work well on our data sets.

}

\section{Related Work}
\label{sec:related}
Copy detection has been studied recently in~\cite{BCM+10, DBH+10a, DBS09a, DBS09b, QAH+13}.
Prior work has focused on effectiveness rather than efficiency of 
detection. As our experiments show, our algorithms
can improve the efficiency over state-of-the-art algorithms
({\sc Pairwise}) by three or more orders of magnitude,
without sacrificing the quality much.

Improving scalability of copy detection has been intensively
studied for text documents and software programs
(surveyed in~\cite{DS11}). For documents, copy detection considers
sharing sufficiently large text fragments as evidence of copying.
The naive strategy looks for the longest common subsequences (LCS),
but can take time $O(n_1 \cdot n_2)$ for documents of sizes
$n_1$ and $n_2$ respectively, and needs to compare every pair of
documents. The first improvement is to build {\em fingerprints} for
each document and only selectively store and compare the fingerprints.
Manber~\cite{M94} fingerprints each
sequence of $Q$ consecutive tokens ($Q$-{\em gram}), and builds
a {\em sketch} with $Q$-grams whose fingerprints are 0 mod $K$;
the space usage is thus only $1 \over K$ of original documents. 
Brin et al.~\cite{BDG95} divides each document
into non-overlapping chunks, where the last unit of each chunk
has a fingerprint that is 0 mod $K$, and sketches each chunk;
again, the space usage is expected to be $1 \over K$ 
of original documents. Schleimer et al.~\cite{SWA03}
also fingerprints each $Q$-gram, but the sketch contains
the smallest fingerprint in each $K$-window; it has the
same space usage but is guaranteed to find reuse of text
with length of at least $K+Q-1$. Another improvement is to
build an index for the sketches, such that two documents are
compared only if they share some fingerprints~\cite{MGS96}.

We also build an inverted index for the provided values
and skip pairs of sources that do not share any value; 
however, our index is different in many ways.
First, each entry in the index is associated with a score,
indicating how strong sharing the value can serve as evidence for copying.
Second, the entries are processed in decreasing order of the scores,
so we consider stronger evidence first and can stop computation
for a pair of sources when we have accumulated sufficient 
evidence for deciding copying or no-copying. Third, source pairs that
share only a few entries with small scores will also be skipped
for copy detection. Finally, we additionally design algorithms
for pruning and incremental copy detection, which have not been discussed
for document copy detection. 
\eat {
Inverted indexes are also used in many other applications:
information retrieval uses inverted index to search keywords~\cite{MRS08};
set similarity join uses inverted index to find overlapping
strings or q-grams~\cite{AGK06, HCK+08, LLL08, SK04}. 
In our context, copy detection requires scanning of the
index and so calls for a wise ordering of the entries. 
}

\eat {
Finally, we show by experiments
that if we apply Fagin's NRA algorithm~\cite{FLN01} for copy detection,
even generating the input to NRA costs more time than that taken by our algorithms.
}

\section{Conclusions}
\label{sec:conclude}
Copy detection has been shown to be crucial for truth finding 
on Web data but meanwhile is a bottle-neck in data fusion.
This paper proposed various methods for improving 
scalability of copy detection on structured data.
Experimental results show that the proposed algorithm 
can reduce copy-detection time by several orders of magnitude
and finish fast on large data sets. 

Our algorithms provide two opportunities for parallelization
in a Hadoop framework. First, when we process each index entry, we can parallelize 
score computation for each pair of sources in that entry. 
Second, we can parallelize computation among entries: whereas parallelizing
on all entries would be hard given the possibly huge number of entries,
{\sc Bound+} provides good insights on which entries can be processed
in parallel. Both approaches are likely to be better
than the strategy that simply extends {\sc Pairwise} by parallelizing copy detection 
for each pair of sources, as the total number of pairs can be huge for big data.
We leave such extensions and an experimental comparison for future work.

{\small
\bibliographystyle{abbrv}
\bibliography{../base}
}
\end{document}